\documentclass[twocolumn]{aastex62}

\hypersetup{linkcolor=red,citecolor=blue,filecolor=cyan,urlcolor=magenta}

\graphicspath{{./}{figures/}}

\shorttitle{The Type I Superluminous Supernova SN\,2017dwh}
\shortauthors{Blanchard et al.}

\begin{document}

\title{A Hydrogen-Poor Superluminous Supernova with Enhanced Iron-Group Absorption: A New Link Between SLSNe and Broad-Lined Type Ic SNe}

%\correspondingauthor{Peter Blanchard}
\email{pblanchard@cfa.harvard.edu}

\author{P.~K.~Blanchard}
\altaffiliation{NSF GRFP Fellow}
\affil{Harvard-Smithsonian Center for Astrophysics, 60 Garden St.~Cambridge, MA 02138, USA}
\author{M.~Nicholl}
\affil{Institute for Astronomy, University of Edinburgh, Royal Observatory, Blackford Hill, Edinburgh, EH9 3HJ, UK}

\author{E.~Berger}
\affil{Harvard-Smithsonian Center for Astrophysics, 60 Garden St.~Cambridge, MA 02138, USA}

\author{R.~Chornock}
\affil{Astrophysical Institute, Department of Physics and Astronomy, 251B Clippinger Lab, Ohio University, Athens, OH 45701, USA}

\author{D.~Milisavljevic}
\affil{Department of Physics and Astronomy, Purdue University, 525 Northwestern Avenue, West Lafayette, IN, 47907, USA}

\author{R.~Margutti}
\affil{Center for Interdisciplinary Exploration and Research in Astrophysics (CIERA) and Department of Physics and Astronomy, Northwestern University, Evanston, IL 60208, USA}

\author{S.~Gomez}
\affil{Harvard-Smithsonian Center for Astrophysics, 60 Garden St.~Cambridge, MA 02138, USA}

\begin{abstract}
We present optical observations of the Type I superluminous supernova (SLSN-I) SN\,2017dwh at $z\!\approx\!0.13$, which reached $M_{i}\!\approx\!-21$ mag at peak.  Spectra taken a few days after peak show an unusual and strong absorption line centered near 3200\,\AA\ that we identify with \ion{Co}{2}, suggesting a high fraction of synthesized $^{56}$Ni in the ejecta.  By $\sim\!1$ month after peak, SN\,2017dwh became much redder than other SLSNe-I, instead strongly resembling broad-lined Type Ic supernovae (Ic-BL SNe) with clear suppression of the flux redward of $\sim\!5000$\,\AA, providing further evidence for a large mass of Fe-group elements.  Late-time upper limits indicate a $^{56}$Ni mass of $\lesssim 0.6$\,M$_\odot$, leaving open the possibility that SN\,2017dwh produced a $^{56}$Ni mass comparable to SN\,1998bw ($\approx\!0.4$\,M$_\odot$). Fitting the light curve with a combined magnetar and $^{56}$Ni model using {\tt MOSFiT}, we find that the light curve can easily accommodate such masses without affecting the inferred magnetar parameters.  We also find that SN\,2017dwh occurred in the least-luminous detected host galaxy to date for a SLSN-I, with $M_{B} = -13.5$ mag and an implied metallicity of $Z\!\sim\!0.08$\,$Z_\odot$.  The spectral properties of SN\,2017dwh provide new evidence linking SLSNe-I with Type Ic-BL SNe, and in particular the high Fe-group abundance may be due to enhanced $^{56}$Ni production or mixing due to asphericity.  Finally, we find that SN\,2017dwh represents the most extreme end of a correlation between continuum shape and \ion{Co}{2} absorption strength in the near-peak spectra of SLSNe-I, indicating that Fe-group abundance likely accounts for some of the variation in their spectral shapes.      
\end{abstract}


\section{Introduction} \label{sec:intro}
The discovery of superluminous supernovae (SLSNe) has transformed our understanding of the diverse ways in which massive stars end their lives.  Type I SLSNe (hereafter SLSNe-I) lack hydrogen in their spectra and are characterized by blue continua and \ion{O}{2} absorption lines before and near peak \citep{Chomiuk2011,Quimby2011,Gal-Yam2012}.  After peak, they eventually evolve to appear similar to normal-luminosity Type Ic SNe {\it at peak} \citep{Pastorello2010}.  

As a result of this spectral similarity, it was natural to consider whether SLSNe-I could be explained as scaled up versions of normal-luminosity Type Ic SNe.  However, the unrealistically large implied nickel fractions of $\gtrsim 50\%$, as well as the mismatch between the nickel mass required by the peak luminosities and the limits imposed by the late-time tail luminosities \citep{Pastorello2010,Inserra2013}, has led to other proposed models. While an overabundant production of $^{56}$Ni in a very massive star through a pair-instability scenario \citep{HegerWoosley2002} has been proposed to explain some SLSNe-I \citep[e.g.~SN\,2007bi;][]{Gal-Yam2009}, no event with both a well-sampled rise and decline has been explained self-consistently by this model \citep[e.g.~][]{Nicholl2013}.

Interaction with a hydrogen-free circumstellar medium \citep{ChevalierIrwin2011} can explain the light curves of SLSNe-I \citep{Chatzopoulos2013,Nicholl2014}, though the lack of narrow and intermediate width emission lines in their spectra is difficult to account for in this scenario.  A magnetar central engine \citep{KasenBildsten2010,Woosley2010}, however, can consistently explain both the light curves \citep{Inserra2013,Nicholl2017} \textit{and} spectra \citep{Dessart2012,Mazzali2016} of SLSNe-I.  Recently, nebular spectra of SLSNe-I have been shown to be similar to the nebular spectra of SNe associated with long-duration gamma-ray bursts \citep[LGRBs;][]{Milisavljevic2013,Nicholl2016,Jerkstrand2016,Jerkstrand2017} and a large sample study of nebular spectra has shown velocity, density, and ionization structures consistent with central engine models \citep{Nicholl2018}. In addition, SLSNe-I occur in host galaxies similar to those of LGRBs \citep{Chen2013,Lunnan2014,Leloudas2015,Perley2016,Schulze2018}.  Finally, the discovery of the luminous SN\,2011kl (though not as luminous as most SLSNe-I) associated with the ultra-long GRB 111209A added yet another connection between SLSNe-I and LGRBs \citep{Greiner2015}.  

SLSNe-I and LGRBs (and their associated SNe) are linked by these observations and by the engine models used to explain them.  But it remains to be understood what factors lead to the formation of SLSNe-I or energetic Type Ic-BL SNe with and without LGRBs.  Theoretical work suggests that a single central engine may explain both jet-powered LGRBs and their associated energetic SNe \citep{Barnes2018} and that adjusting the engine parameters may lead to the formation of SLSNe-I \citep{Metzger2015,Margalit2018}.  It also may be the case that the engines are different \citep[e.g., magnetars for SLSNe-I and black holes for LGRBs;][]{Woosley1993,MacFadyen1999}.  

Despite the expected similar progenitors, there are few transitional events.  LGRB engines release the bulk of their energy on a timescale of seconds to minutes and their associated SNe (hereafter GRB-SNe) produce a large nickel mass of a few tenths of solar masses, whereas SLSN-I engines release energy on a timescale of days to weeks, thereby enabling a significant boost in the SN optical luminosity. The connections between these classes raises the question of whether there exist hybrid events with LGRB-like nickel masses and/or mixing, but with SLSN-like luminosities.

Here we present observations of SN\,2017dwh, a SLSN-I with evidence for a significantly higher nickel fraction than seen in previous SLSNe-I.  In particular, its spectrum shows significant absorption from iron-peak elements and a strong match to the red colors of Type Ic-BL SNe at phases greater than a few weeks after peak brightness.  Our observations show SN\,2017dwh is an unprecedented event with transitional properties between SLSNe-I and GRB-SNe.

The paper is structured as follows.  In Section 2 we present the identification of SN\,2017dwh as a SLSN-I.  In Section 3 we present spectroscopic observations of SN\,2017dwh and comparisons to other SNe.  In Section 4 we analyze and compare the light curve.  In Section 5 we discuss the key results and implications for our understanding of the connection between SLSNe-I and their lower luminosity counterparts, and we conclude in Section 6.

In this paper we use $H_{0} = 67$ km s$^{-1}$ Mpc$^{-1}$, $\Omega_{m} = 0.32$, and $\Omega_{\Lambda} = 0.68$ \citep{Planck2013}, resulting in a luminosity distance of 636.4 Mpc to SN\,2017dwh (for $z=0.13$ determined from spectroscopic comparisons).  The Galactic extinction along the line of sight to SN\,2017dwh is $E(B-V) = 0.0123 \pm 0.0004$ mag \citep{SF2011}. 

\begin{figure*}
\centering
\includegraphics[scale=0.52]{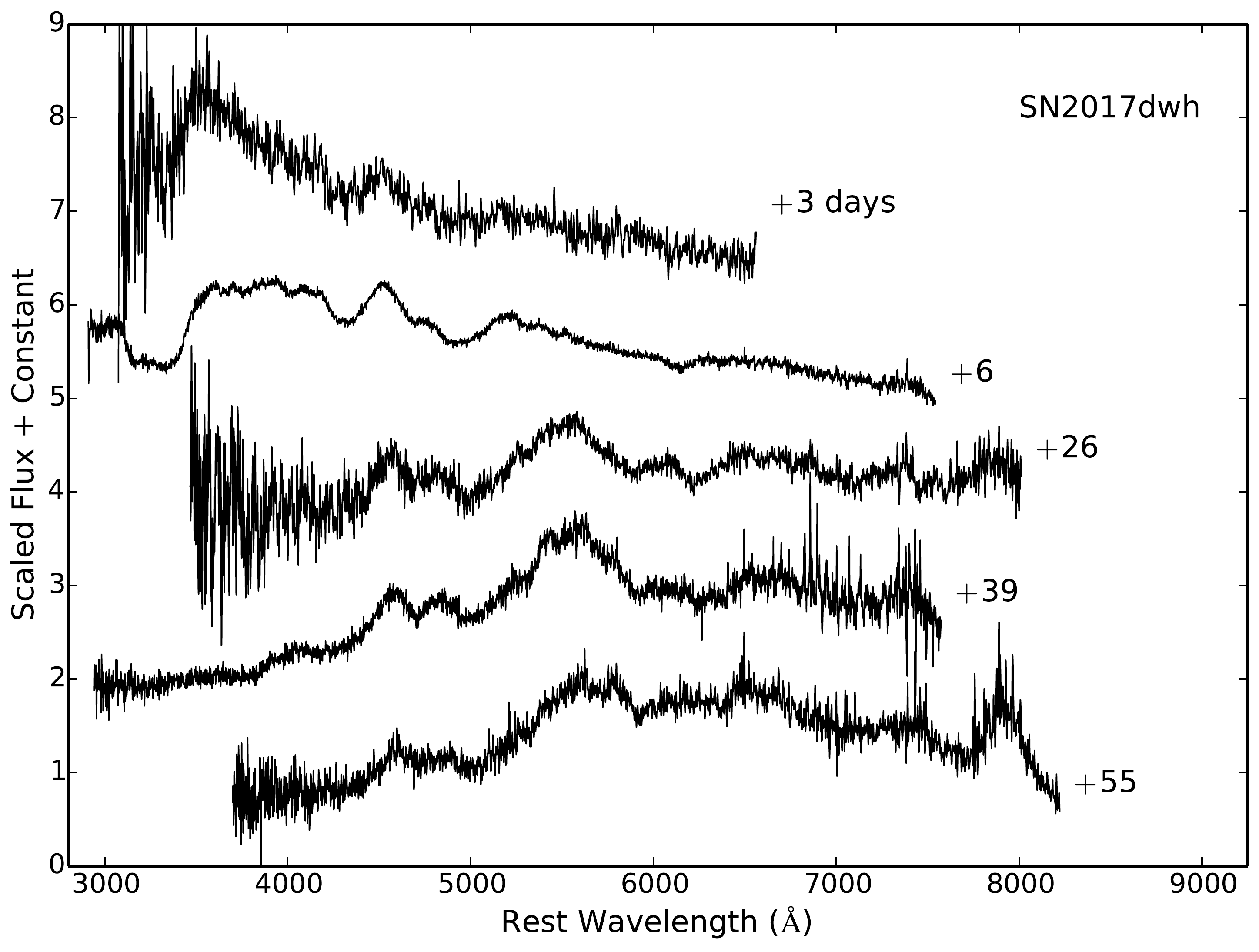}
\caption{Optical spectra of SN\,2017dwh.  The rest-frame phases relative to the observed $i$-band maximum are shown.  In the first two spectra, SN\,2017dwh exhibits a notably strong absorption feature centered near 3200 \AA, which is not common in other SLSNe-I.  The SN subsequently evolves rapidly to a redder continuum.}
\label{fig:spec}
\end{figure*}

\begin{figure*}
\centering
\includegraphics[scale=0.415]{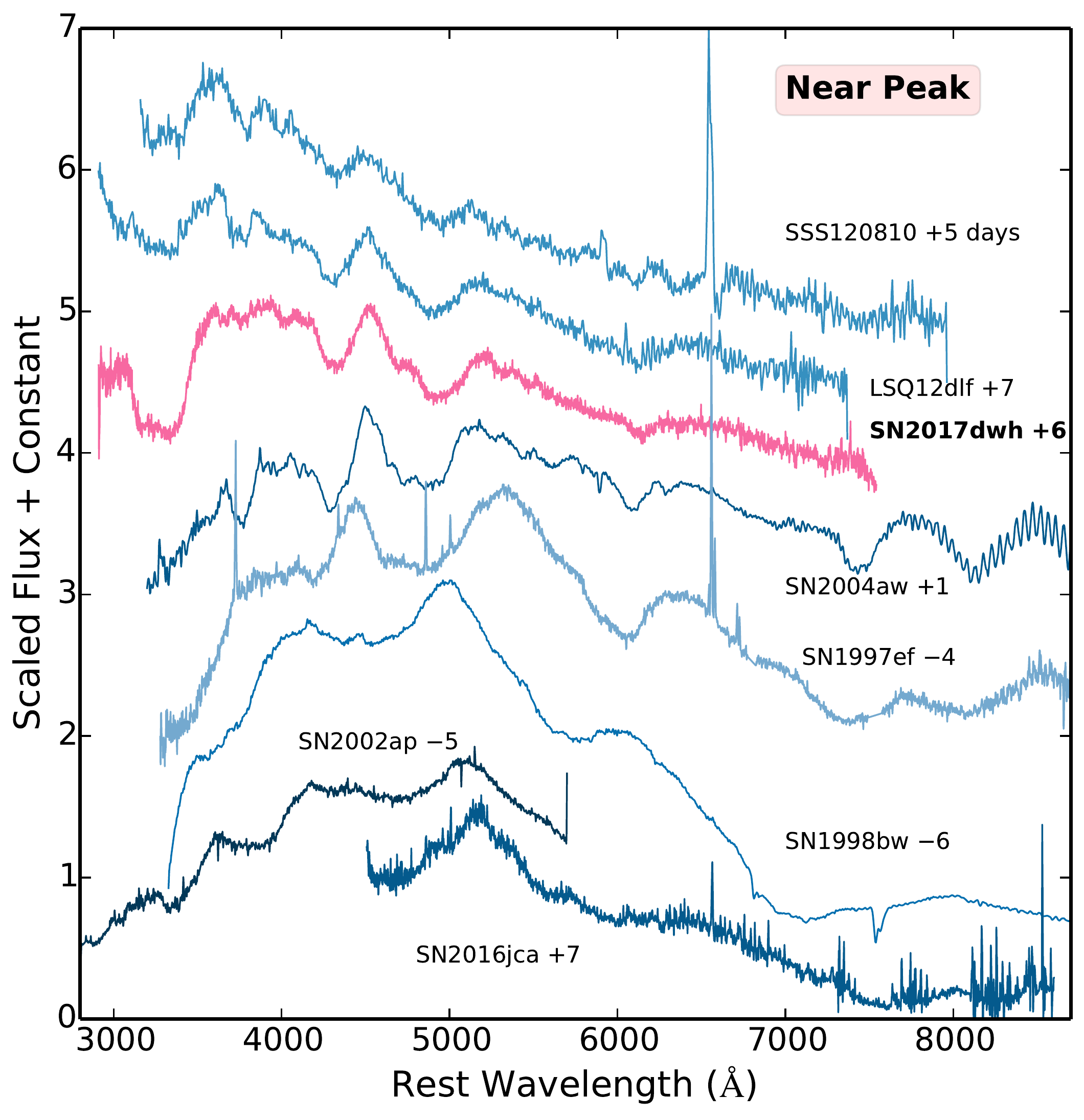}
\includegraphics[scale=0.415]{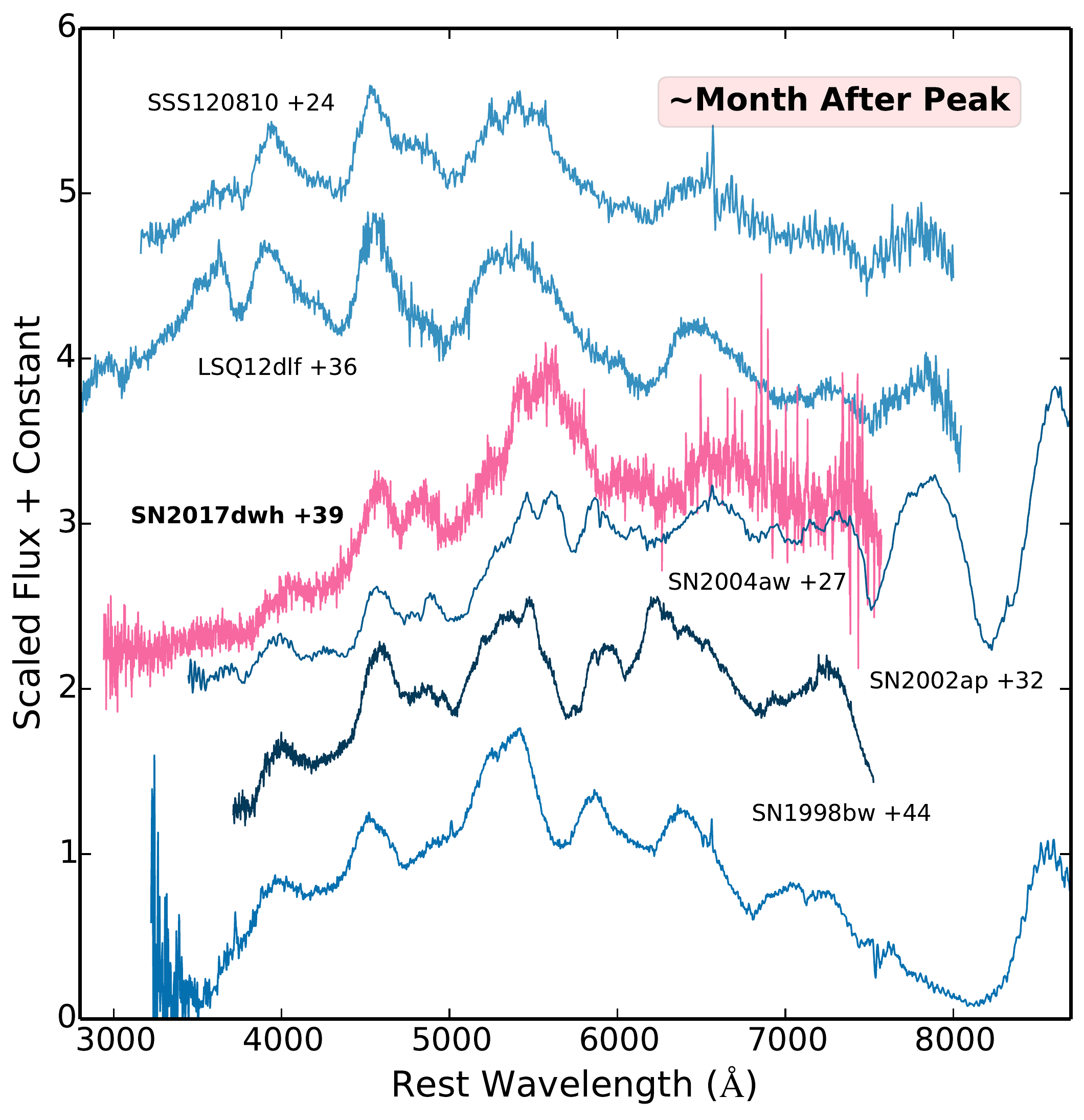}
\caption{Spectral comparisons of SN\,2017dwh near peak (Left) and about a month after peak (Right) with the SLSNe-I SSS120810 and LSQ12dlf, the Type Ic SN\,2004aw, the Type Ic-BL SNe 1997ef and 2002ap, and the GRB-SNe 1998bw and 2016jca.  Near peak, SN\,2017dwh closely matches SLSNe-I at wavelengths redward of $\approx\!4000$ \AA.  A notable difference is the much stronger absorption feature near 3200 \AA\ in SN\,2017dwh.  SN\,2017dwh does not exhibit velocities as high as those in Type Ic-BL SNe with and without GRBs, although several of these events show evidence for similarly strong absorption blueward of $\approx\!3500$ \AA.  About a month after peak, SN\,2017dwh evolved to closely match Type Ic/Ic-BL SNe at similar phases, while diverging considerably from the typical evolution of SLSNe-I which stay hotter for longer.}
\label{fig:speccomp}
\end{figure*}

\section{Identification of SN\,2017dwh as a SLSN-I} \label{sec:obs}

SN\,2017dwh was first detected by the Catalina Real-Time Transient Survey \citep[CRTS;][]{Drake2009} on 22 April 2017 with a magnitude of $m_{\rm CSS} = 20.39 \pm 0.21$ and designated CSS170425:143443+312917.  A subsequent CRTS detection on 25 April 2017 with a magnitude of $m_{\rm CSS} = 19.12 \pm 0.08$ indicated a rising transient.  SN\,2017dwh was also detected by the Asteroid Terrestrial-impact Last Alert System \citep[ATLAS;][]{Tonry2018} on 26 April 2017 (designated ATLAS17fau) with a magnitude of $m = 18.73 \pm 0.11$ in the cyan filter and by the Pan-STARRS Search for Transients \citep[PSST;][]{Huber2015} on 11 May 2017 (designated PS17dbf) with a magnitude of $m_{i} = 18.01 \pm 0.04$.  Inspecting the Pan-STARRS 3$\pi$ stacked images we found no detection of a host galaxy in any filter at the position of SN\,2017dwh (upper limit in $i$-band of $m_i > 23.4$), leading us to select the object for spectroscopic follow-up.  We obtained a spectrum of the event on 19 May 2017, which showed features matching \ion{Fe}{2}+\ion{Mg}{2} absorption lines at $4200-5200$ \AA, as seen shortly after peak in SLSNe-I \citep{Pastorello2010,Gal-Yam2012}. The inferred redshift of $z \approx 0.13$ implied an absolute magnitude of $M_{i} \approx -20.9$ for the PSST detection, confirming SN\,2017dwh as a SLSN-I.  However, the spectrum also exhibited an unusual absorption line near 3200 \AA, motivating continued follow-up.

\section{The Unusual Spectra of SN\,2017dwh}
\label{sec:spec}
We obtained five epochs of spectroscopy of SN\,2017dwh using the FAST spectrograph \citep{Fabricant1998} on the 60-inch telescope at Fred Lawrence Whipple Observatory (FLWO), the Blue Channel spectrograph \citep{Schmidt1989} on the 6.5-m MMT telescope, the Ohio State Multiple Object Spectrograph \citep[OSMOS;][]{Martini2011} on the 2.4-m Hiltner telescope at MDM Observatory, and the Inamori-Magellan Areal Camera and Spectrograph \citep[IMACS;][]{Dressler2011} on the 6.5-m Magellan Baade telescope.  Details of the observations are presented in the Appendix.  We extracted 1D wavelength-calibrated spectra using IRAF and used observations of standard stars obtained on the same nights for relative flux calibration.  We correct the spectra for Galactic extinction and transform them to the rest-frame of SN\,2017dwh.  None of the spectra reveal host galaxy emission lines, and so we infer the redshift of $z = 0.13$ from matching spectral features in other SLSNe-I and Type Ic SNe. 

In Figure \ref{fig:spec} we show the spectra of SN\,2017dwh spanning $+3$ to $+55$ rest-frame days after the observed maximum in $i$-band.  While fairly noisy, the first spectrum at $+3$ days exhibits a blue continuum, several absorption features at $4200-5200$ \AA, and evidence for significant absorption blueward of $\approx\!3500$ \AA.  The higher signal-to-noise ratio spectrum at $+6$ days clearly reveals the same absorption features, showing that the blue absorption corresponds to a strong line centered at $\approx\!3200$ \AA.  

The spectrum of SN\,2017dwh then undergoes a significant transition in the next 20 days.  In particular, the spectrum at $+26$ days exhibits a much redder continuum and additional absorption features at $5500-6500$ \AA, creating a broad spectral peak near 5500 \AA.  The significant spectral transition between the spectrum at $+6$ and $+26$ days after peak is likely a combination of a rapidly cooling continuum and increased line absorption.  The same general shape and features remain present in the spectra obtained at $+39$ and $+55$ days, with the additional development of a P-Cygni feature near 7800 \AA\ in the $+55$ day spectrum.

\begin{figure*}[t]
\centering
\includegraphics[scale=0.41]{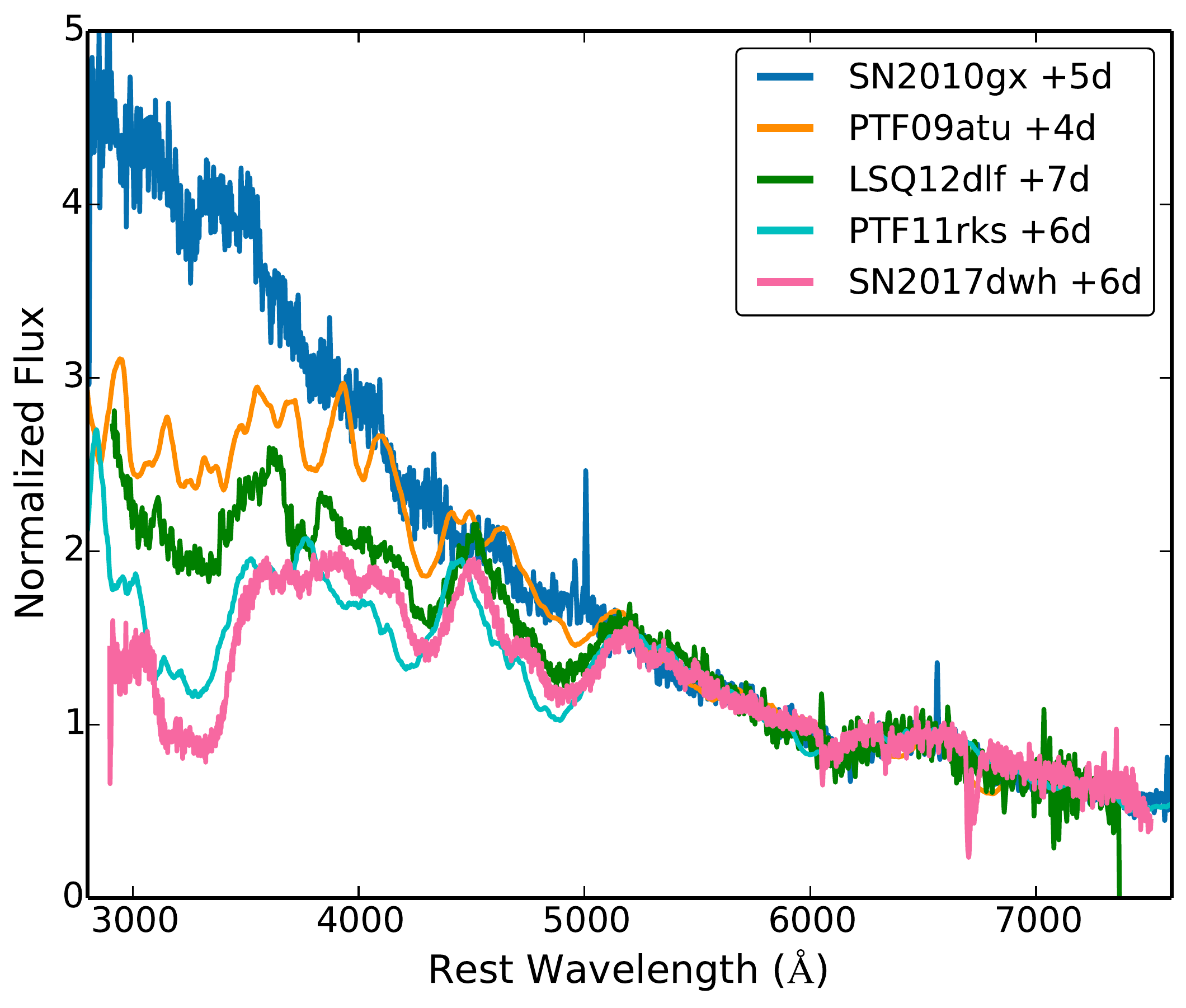}
\includegraphics[scale=0.41]{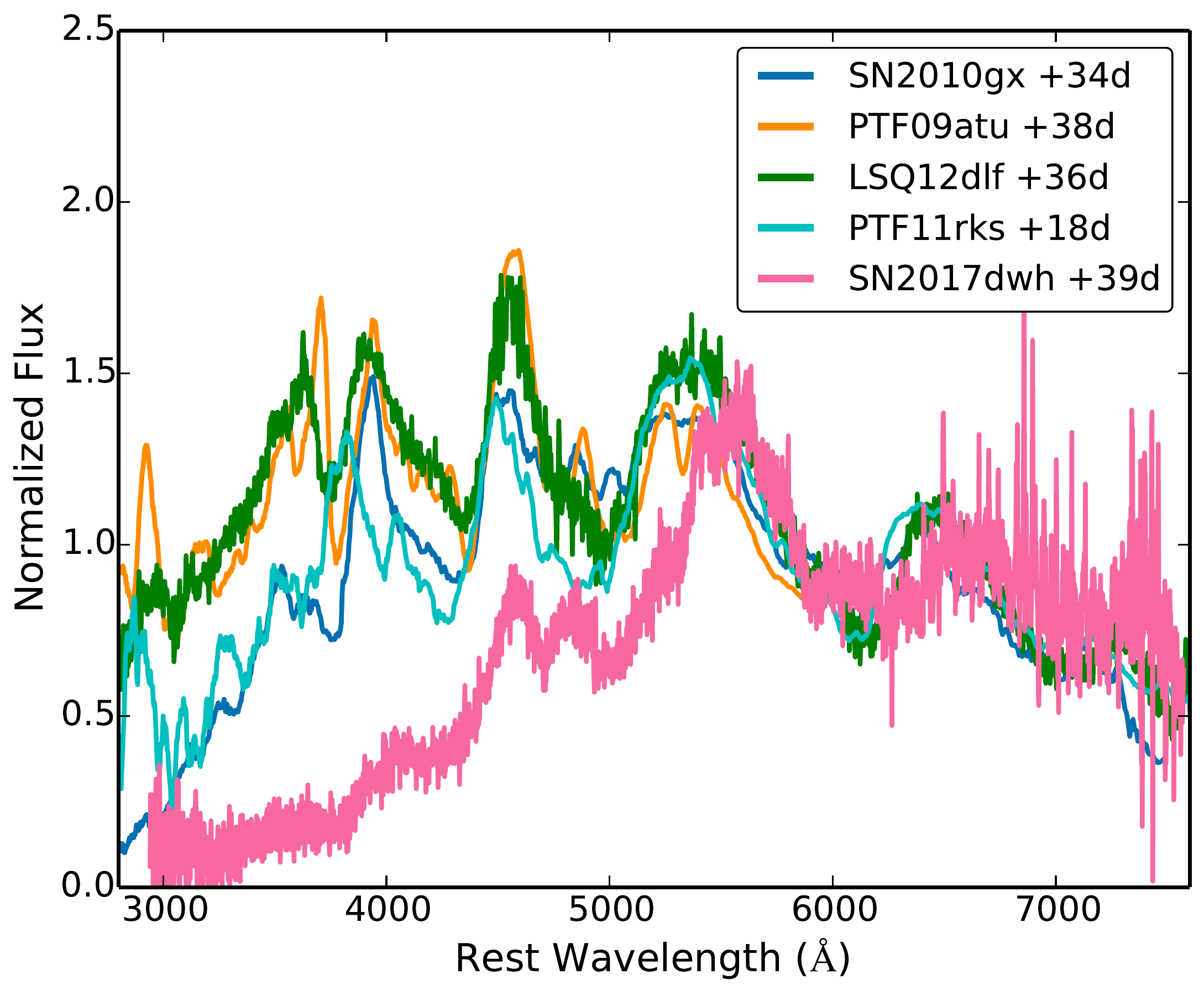}
\caption{Left: Near peak spectra of SN\,2017dwh compared to SN\,2010gx, PTF09atu, LSQ12dlf, and PTF11rks (all normalized around 6000 \AA) showing a range of absorption strengths near 3200 \AA\ with SN\,2017dwh exhibiting the most extreme absorption.  The absorption strength appears to correlate with continuum shape, with the bluest events exhibiting the weakest absorption line.  Right: Spectra of the same objects about a month after peak showing that SN\,2017dwh transitions to a much redder spectrum than the other SLSNe-I.}
\label{fig:SLSNcomp}
\end{figure*}

\begin{figure*}
\centering
\includegraphics[scale=0.5]{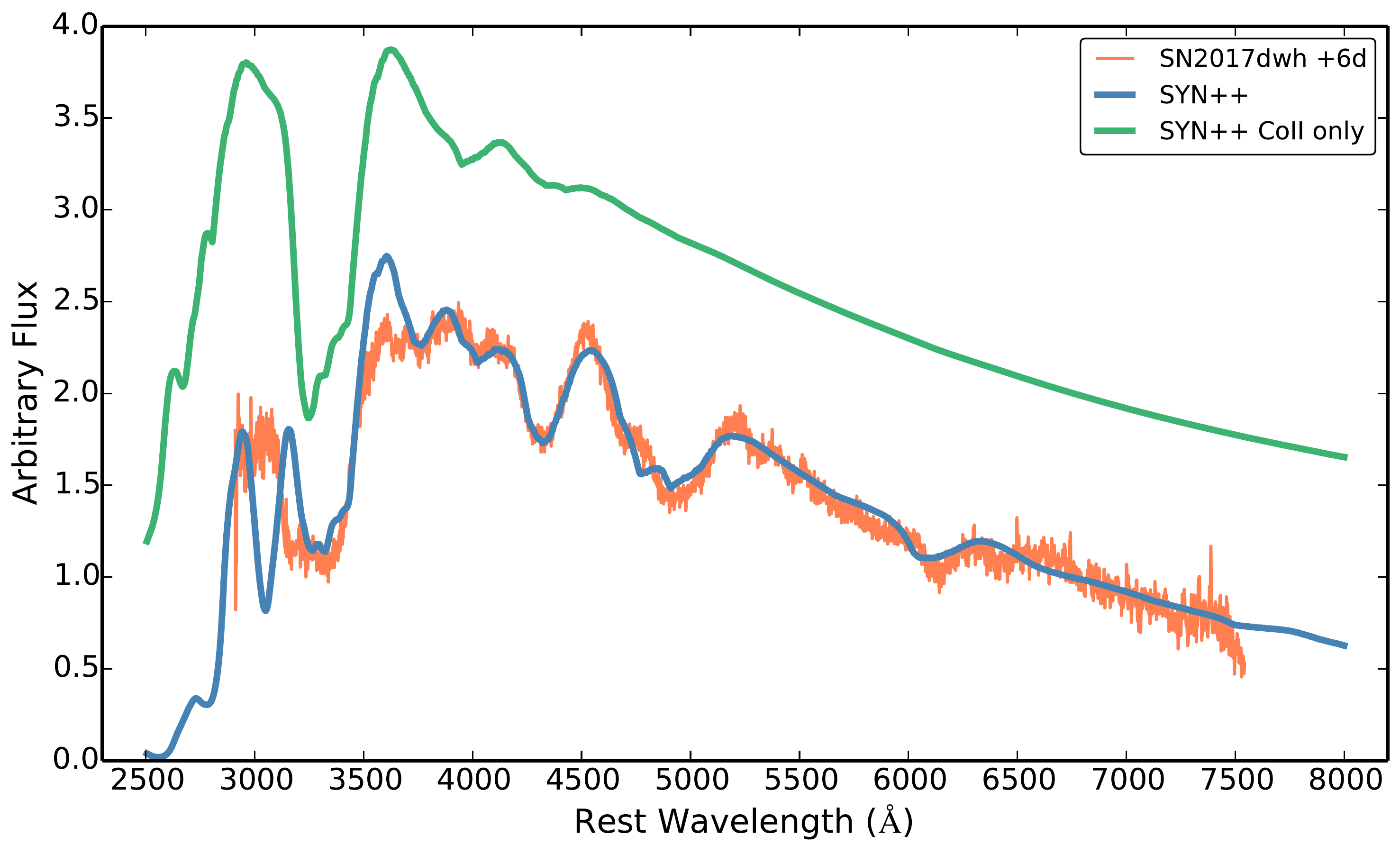}
\caption{Spectrum of SN\,2017dwh at 6 rest-frame days after peak, compared to synthetic spectra generated with {\tt SYN++}.  We show a synthetic spectrum with a composition consisting of \ion{Fe}{2}, \ion{Mg}{2}, \ion{Si}{2}, \ion{Ca}{2}, and \ion{Co}{2}, as well as a spectrum consisting of only \ion{Co}{2}.  The synthetic spectra have underlying blackbody continuum temperatures of 8500 K and photospheric velocities of 16,000 km s$^{-1}$.  We find that most of the features redward of 3500 \AA\ can be explained by \ion{Fe}{2}, \ion{Mg}{2}, \ion{Si}{2}, and \ion{Ca}{2}, as observed in previous SLSNe-I.  In addition, we identify the strong absorption centered near 3200 \AA\ as likely resulting from a large contribution from \ion{Co}{2}.  However, \ion{Co}{2} does not account for the full width of the line, suggesting other ions may be contributing.}
\label{fig:synow}
\end{figure*}

\subsection{Spectral Comparisons}
In Figure \ref{fig:speccomp} we compare the spectra of SN\,2017dwh to those of other SLSNe-I and Type Ic and Ic-BL SNe.  SN\,2017dwh initially shows similar absorption features to those seen in the SLSNe-I LSQ12dlf and SSS120810 \citep{Nicholl2014} at similar phases.  The lines at $4200-5200$ \AA\ are usually attributed to \ion{Fe}{2} and \ion{Mg}{2} \citep{Pastorello2010,Gal-Yam2012,Inserra2013}.  Many SLSNe-I still show \ion{O}{2} lines at this phase, so the earlier appearance of \ion{Fe}{2} and \ion{Mg}{2} in these events is likely due to having a lower photospheric temperature at peak.   SN\,2017dwh does not show as strong an absorption line at 3800 \AA, usually attributed to \ion{Ca}{2}, as LSQ12dlf and SSS120810.  While LSQ12dlf and SSS120810 show evidence for absorption blueward of $\approx\!3500$ \AA, SN\,2017dwh exhibits a significantly stronger feature centered near 3200 \AA.  

SN\,2017dwh appears to exhibit a much cooler continuum than these other SLSNe-I, further setting it apart from the typical spectral characteristics of SLSNe-I. Increased absorption may also be playing a role in the blue portion of the spectrum due to metal line blanketing.  Even so, the continuum at this phase is still clearly bluer than normal-luminosity Type Ic SNe, such as SN\,2004aw \citep[][also shown in Figure \ref{fig:speccomp}]{Taubenberger2006}.  Therefore, SN\,2017dwh at peak appears to exhibit an intermediate color temperature between normal Type Ic SNe and typical SLSNe-I.

One defining characteristic of the spectroscopic evolution of SLSNe-I is that their continua remain hotter for longer than their lower luminosity counterparts.  After peak, SN\,2017dwh continues to diverge from this typical SLSN-I evolution by rapidly transitioning to a red spectrum.  Compared to SSS120810 and LSQ12dlf, SN\,2017dwh has cooled significantly by about a month after peak and instead has a much better match to the shape of SN\,2004aw.  We broaden the comparison by also examining the large sample of SLSN-I spectra released by \citet{Quimby2018}.  We find a few objects in their sample with absorption near 3200 \AA\ (e.g., PTF11rks and PTF09atu).  In Figure \ref{fig:SLSNcomp} we compare SN\,2017dwh to PTF11rks, PTF09atu, SN\,2010gx \citep{Pastorello2010}, and LSQ12dlf after normalizing the spectra by their average flux near 6000 \AA.  These events show a range of continuum shapes and absorption strengths near peak with the bluest object, SN\,2010gx, exhibiting the weakest absorption, and the reddest object, SN\,2017dwh, exhibiting the strongest absorption. While SN\,2017dwh shows the strongest absorption near 3200 \AA, PTF11rks is a close analog.  After peak, SN\,2017dwh shows the most extreme evolution among these SLSNe-I to a much redder spectrum.  

We now turn to a comparison with the spectra of the Type Ic-BL SNe 1997ef and 2002ap \citep[not accompanied by LGRBs;][]{Iwamoto2000,Gal-Yam2002,Mazzali2002,Foley2003} and the Type Ic-BL SNe 1998bw and 2016jca\footnote{We observed SN\,2016jca with GMOS on Gemini-South under program GS-2016B-DD-9.} \citep[accompanied by LGRBs;][]{Patat2001,Ashall2017,Cano2017} shown in Figure \ref{fig:speccomp}.  SN\,2017dwh clearly does not show significantly broadened and blueshifted lines near peak like Type Ic-BL SNe.  SN\,2017dwh's spectrum near peak is much more similar to SLSNe-I and SN\,2004aw.    However, the $+26$, $+39$, and $+55$ day spectra of SN\,2017dwh bear strong resemblance to the spectra of Type Ic-BL SNe at similar phases.  In the right panel of Figure \ref{fig:speccomp}, we show the $+39$ day spectrum of SN\,2017dwh compared to SN\,2002ap at $+32$ days and SN\,1998bw at $+44$ days.  SN\,2017dwh shows the same spectral shape with similar spectral features. The exception is an absorption near 5800\,\AA\ that is not seen in SN\,2017dwh, generally attributed to \ion{Na}{1} D in Type Ic-BL SNe.  We also note that the maximum-light spectra of Type Ic-BL SNe, as with SN\,2017dwh, show significant absorption blueward of $\approx\!3500$ \AA\ (though in SN\,2002ap it is not particularly prominent).  

In summary, we find that SN\,2017dwh exhibits stronger absorption near 3200 \AA\ and a cooler continuum than typical SLSNe-I near peak, and then rapidly evolves to match the red spectral shapes of Type Ic/Ic-BL SNe about a month after peak.  These characteristics set SN\,2017dwh apart from other SLSNe-I, despite its similar peak luminosity.

\begin{figure*}
\centering
\includegraphics[scale=0.65]{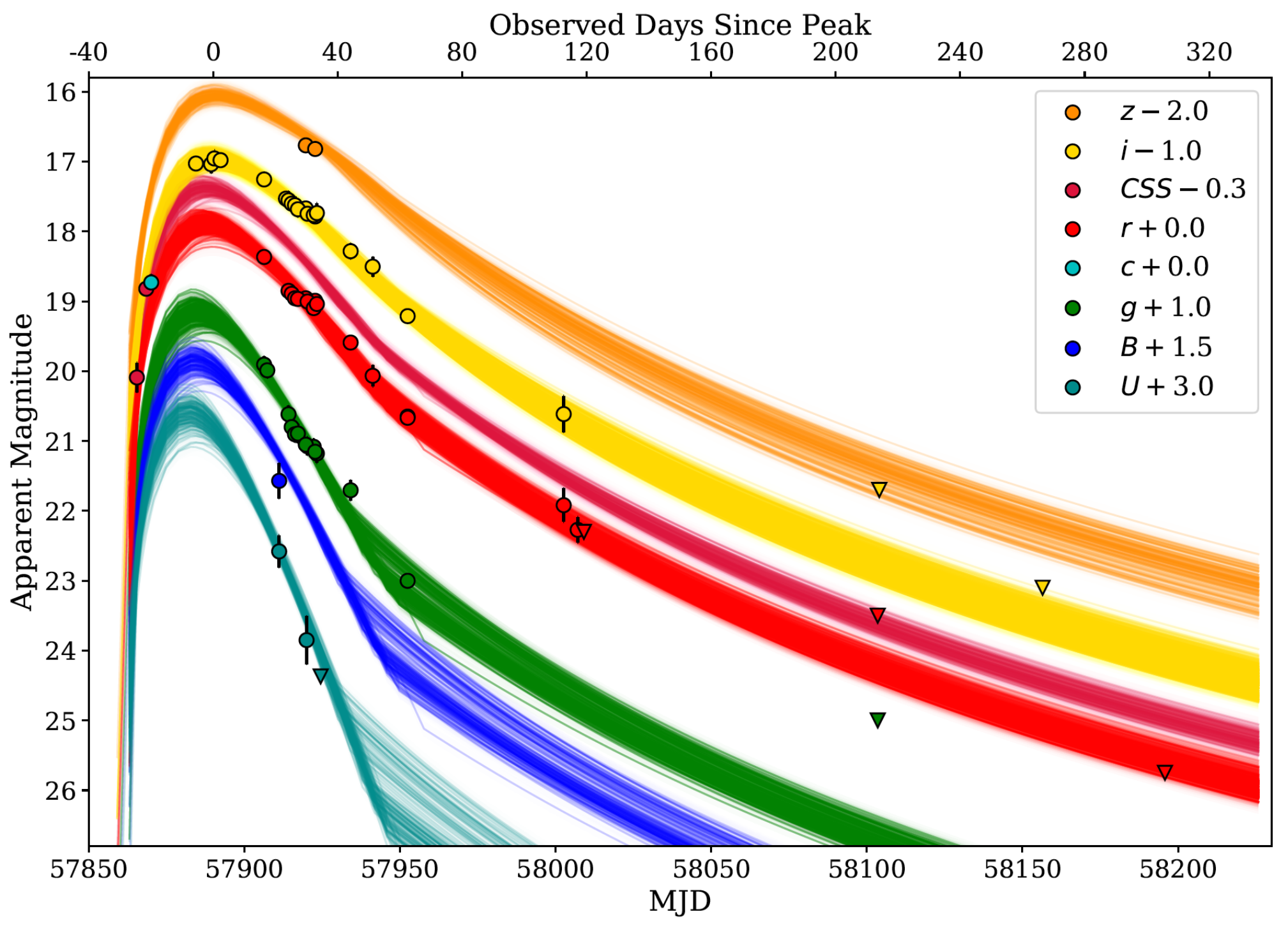}
\caption{Light curves of SN\,2017dwh including our ground-based and \textit{Swift}/UVOT observations and data from CRTS, ATLAS, and PSST.  We also show an ensemble of magnetar model realizations from {\tt MOSFiT}.  The corresponding posterior parameter distributions are shown in Figure \ref{fig:corner} and the median parameter values are listed in Table \ref{tab:param}.  The models shown here are for the magnetar-only fit.  We also perform a fit with a combined magnetar and $^{56}$Ni model and find similar results, indicating the effect of $^{56}$Ni input can be ``hidden'' in the light curve due to the dominating effect of the magnetar energy input.}
\label{fig:mosfit}
\end{figure*}

\subsection{Identification of the 3200\,\AA\ line as \ion{Co}{2}}

The strong absorption near 3200 \AA\ in SN\,2017dwh, and its weaker presence in a few other SLSNe-I (Figure \ref{fig:SLSNcomp}), raises the question of its origin.  Such strong lines in SLSNe-I are typically only seen further into the UV \citep{Mazzali2016,Yan2017}.  However, strong absorption in this region has been observed in some peculiar Type Ia SNe.  The well-studied overluminous Type Ia SN 1991T \citep{Filippenko1992} exhibited strong absorption near 3200 \AA. Through spectral modeling, this was attributed to \ion{Co}{3} and \ion{Fe}{3} at pre-peak phases, with a progressively larger contribution from \ion{Co}{2} near and after peak \citep{Sasdelli2014}.  While the similarity of this line to those seen in SN\,1991T suggests an identification with Fe-group elements like cobalt, we note that SN\,2017dwh is clearly not a Type Ia SN given its resemblance to SLSNe-I near peak and to Type Ic and Ic-BL SNe a month after peak.  In addition, SN\,2017dwh is much more luminous than any Type Ia SN.

Synthetic spectral modeling has been successfully used to identify some of the unusual spectral features seen in SLSNe-I such as the \ion{O}{2} absorption features present in early spectra \citep{Quimby2011}.  We use {\tt SYN++} \citep{Thomas2011} to explore possible identifications of the strong absorption line near 3200 \AA.  In Figure \ref{fig:synow} we show the $+6$ day spectrum compared to a synthetic spectrum produced with a composition consisting of \ion{Fe}{2}, \ion{Mg}{2}, \ion{Si}{2}, \ion{Ca}{2}, and \ion{Co}{2}.  We also show a synthetic spectrum produced using only \ion{Co}{2}.  The synthetic spectra have temperatures of 8500 K for the underlying blackbody continuum and photospheric velocities of 16,000 km s$^{-1}$.  We find that the red side of the absorption line at 3200 \AA\ can be well matched by \ion{Co}{2}.  

However, the observed width of the feature is wider than the other lines in the synthetic spectrum.  Given the large number of lines from Fe-group elements in the near-UV \citep[models of Type Ia and Ic/Ic-BL SNe show considerable flux suppression below 4000 \AA;][]{Mazzali2000,Mazzali2002,Mazzali2017}, it is likely that other ions are also contributing (perhaps \ion{Co}{3} and \ion{Fe}{3} as in SN\,1991T).  The line width may also be explained if the cobalt extends to higher velocity zones due to enhanced outward mixing.  

\begin{figure*}
\centering
\includegraphics[scale=0.4]{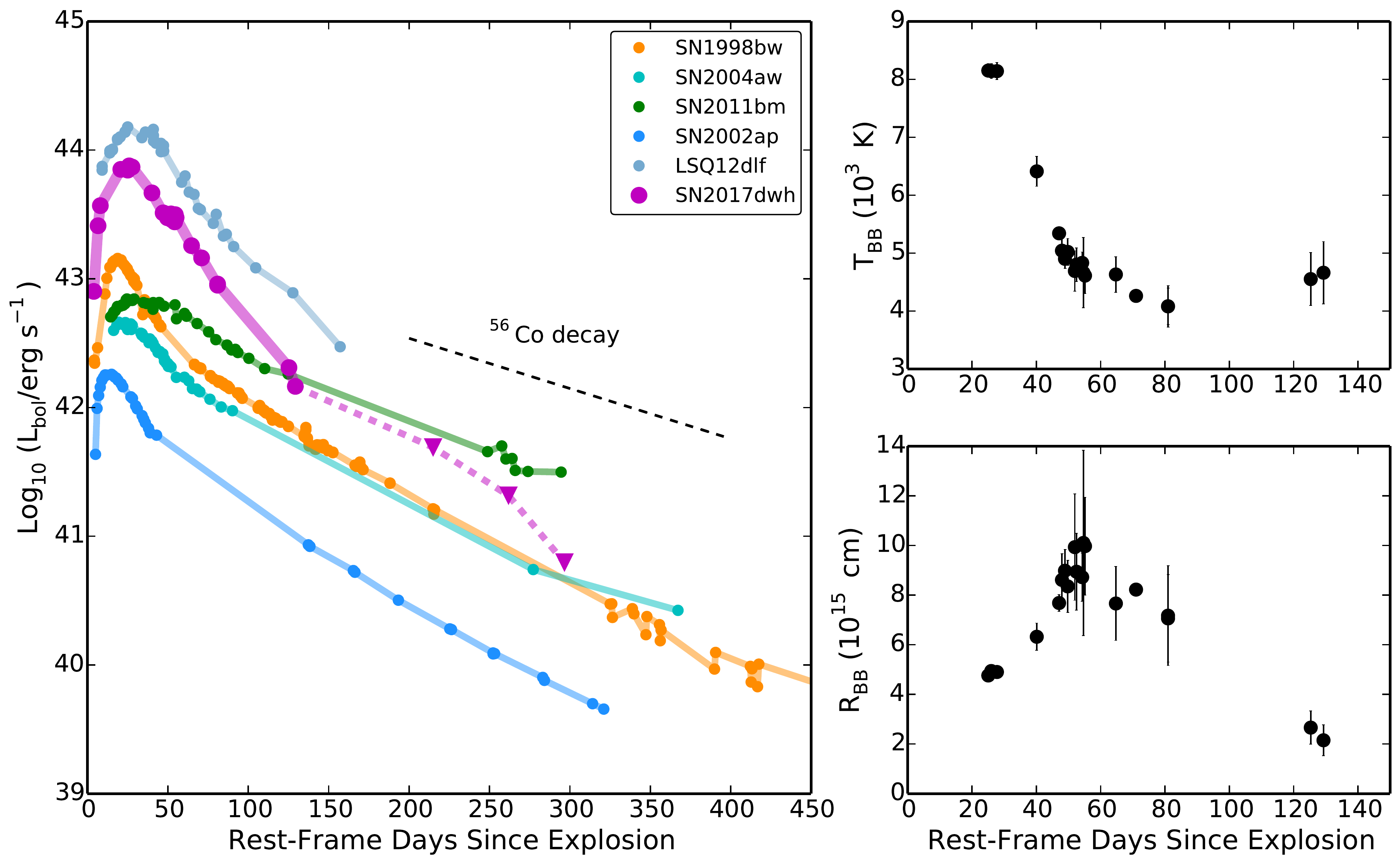}
\caption{Left: Bolometric light curve of SN\,2017dwh compared to those of SN\,1998bw, SN\,2002ap, SN\,2004aw, SN\,2011bm, and LSQ12dlf.  The upper limits on the luminosity of SN\,2017dwh at times $\gtrsim200$ rest-frame days since explosion (connected to the early light curve by a dashed line) show that the SN dropped below the tail luminosity of the Type Ic SN\,2011bm, suggesting SN\,2017dwh did not produce as much $^{56}$Ni as SN\,2011bm ($\approx\!0.7$ M$_{\odot}$).  We are unable to rule out a late-time luminosity as bright as SN\,1998bw or SN\,2004aw, which had $^{56}$Ni masses of $\approx 0.4$ M$_{\odot}$ and $\approx 0.2$ M$_{\odot}$, respectively.  The upper limits imply a limit on the nickel mass produced in SN\,2017dwh of $M_{\rm Ni}$ $\lesssim0.6$ M$_{\odot}$.  Top right: Blackbody temperature evolution showing the rapid cooling.  Bottom right: Corresponding evolution of the inferred photospheric radius.}
\label{fig:LCcomp}
\end{figure*}

\section{Light Curve of SN\,2017dwh} 
\label{sec:LC}
We obtained images of SN\,2017dwh using the 48-inch telescope at FLWO in the $gri$ filters.  We also obtained $gri$ imaging using IMACS on the 6.5-m Magellan Baade telescope and the Low Dispersion Survey Spectrograph \citep[LDSS3c;][]{Stevenson2016} on the 6.5-m Magellan Clay telescope, $griz$ imaging with the 1.3-m McGraw-Hill telescope at MDM Observatory, and an $i$-band image using Binospec on the 6.5-m MMT telescope.  In addition, we triggered observations with the Neil Gehrels \textit{Swift} satellite, obtaining one epoch of imaging with the UV/Optical Telescope \citep[UVOT;][]{Roming05} in the $uvw2$, $uvm2$, $uvw1$, $u$, $b$, and $v$ filters and four additional epochs in $u$-band.

Photometry on the ground-based images was performed using PSF fitting implemented by the IRAF {\tt daophot} package.  Instrumental magnitudes were calibrated to AB magnitudes on the Pan-STARRS 3$\pi$ photometric system using comparison stars in the field.  \textit{Swift}/UVOT photometry was performed using aperture photometry and calibrated to the Vega system using the prescription of \citet{Brown09} and the updated calibration files and zeropoints from \citet{Breeveld11}.  The photometry is listed in the Appendix.

In Figure \ref{fig:mosfit} we show the resulting light curves of SN\,2017dwh, including data from CRTS, ATLAS, and PSST.  The CRTS data points are averages of several individual observations from the same nights.  Both CRTS and ATLAS captured the rising portion of the light curve, while our observations began at peak.  We adopt MJD 57890 as the time of peak, measured from the $i$-band maximum. A comparison of the timescales to other SLSNe-I indicates that SN\,2017dwh is a typical fast evolving event, with a $\approx\!19$ day rise time and a rapid $\approx\!28$ day decline time, defined as the time to rise or decline by a factor of $e$ below the peak brightness \citep{Nicholl2015}.  SN\,2017dwh exhibits a progressively faster decline in bluer bands, decreasing by $\approx$ 1.5 mags in $u$-band over $\approx$ 10 days, after which it was no longer detected.  

In the optical bands we detected SN\,2017dwh to about 117 days after peak, when it had $r\approx22$ mag and subsequently became Sun-constrained.  When the SN was visible again at about 213 days after peak, it was no longer detected, to limits of $m_{g} > 24.0$, $m_{r} > 23.5$, and $m_{i} > 22.5$ mag.  At 267 days after peak, we obtained an additional deep upper limit of $m_{i} > 24$ mag.  At 306 days after peak we obtained deeper images using LDSS3c and detected a source at the position of SN\,2017dwh with $m_{g} = 25.52 \pm 0.11$ and $m_{r} = 25.54 \pm 0.15$.  While the source is unresolved, the $g-r$ color suggests the source is more likely the host galaxy than the SN.  At 380 days after peak, we obtained another epoch of deep $g$-band imaging.  Using {\tt HOTPANTS} \citep{Becker2015} we subtracted this image from the $g$-band image at 306 days after peak and found no residual flux, suggesting a constant flux between the two epochs, consistent with the source being the host galaxy.  At 410 days after peak we obtained additional deep $r$-band and $i$-band imaging.  Performing image subtraction with the $r$-band image from 306 days after peak, we also found no residual flux.  We measure a 3$\sigma$ upper limit of $m_{r} > 25.7$ on the flux of SN\,2017dwh at 306 days by injecting point sources at the position of SN\,2017dwh and determining at what magnitude a source is recovered at 3$\sigma$ significance in the subtracted image.        

In Figure \ref{fig:LCcomp} we show the rest-frame bolometric light curve of SN\,2017dwh calculated by integrating the observed flux and fitting a blackbody to the $gri$ light curves to estimate the flux outside the observed bands.  We assume negligible internal host galaxy reddening, consistent with the low-luminosity host galaxy (see Section \ref{sec:host}).  To determine the bolometric output at peak brightness, where only $i$-band data exists, we use our spectra taken a few days after peak (see Section \ref{sec:spec}) to calculate synthetic $g$ and $r$ magnitudes at peak.  The pre-peak bolometric luminosity was then determined by assuming the same temperature as at peak.  We find that at peak, SN\,2017dwh exhibited a photospheric temperature of $\approx8000$ K but we caution that a blackbody is a crude approximation when there is strong absorption.  The SN then rapidly cools to $\approx4500$ K in about 25 rest-frame days and remains roughly at this temperature for at least the next 75 days.  The inferred photospheric radius starts near $5 \times 10^{15}$ cm and reaches $\approx9 \times 10^{15}$ cm in $\approx25$ rest-frame days, suggesting a photospheric velocity of $\approx18,000$ km s$^{-1}$.  As seen in previous SLSNe-I and many other SNe, the photosphere then begins to recede into the ejecta, reaching $2 \times 10^{15}$ cm after another $\approx75$ days.         

\subsection{Light Curve Comparisons}
Due to the strong absorption from Fe-group elements in the spectra, we examine the light curve for additional evidence of radioactive decay.  In Figure \ref{fig:LCcomp} we compare the bolometric light curve of SN\,2017dwh with those of the SLSN-I LSQ12dlf and several Type Ic and Ic-BL SNe.  We calculated comparison bolometric light curves from $UBVRI$ photometry of SN\,1998bw \citep{Galama1998,McKenzie1999,Sollerman2002}, SN\,2004aw \citep{Taubenberger2006}, SN\,2011bm \citep{Valenti2012}, and SN\,2002ap \citep{Foley2003} retrieved from the Open Supernova Catalog \citep{Guillochon2017}.  The bolometric light curve of LSQ12dlf was calculated using $BVRI$ photometry from \citet{Nicholl2014}.  In each case, we use a blackbody fit to estimate the flux outside these observed bands.  Therefore the curves in Figure \ref{fig:LCcomp} represent estimates of the true bolometric luminosity.  While a simple blackbody fit may not account for all of the unobserved flux, it provides sufficient accuracy for the purpose of comparison.  SN\,2017dwh's peak luminosity is about an order of magnitude larger than the comparison Type Ic and Ic-BL events.  

Comparing the upper limits at $>\!200$ rest-frame days after explosion (explosion epoch estimated from our modeling in Section \ref{modeling}) to the cobalt-tail luminosities of various Type Ic and Ic-BL SNe we can make inferences about the relative nickel mass produced in SN\,2017dwh.  We find that SN\,2017dwh drops below the cobalt-tail luminosity of SN\,2011bm by about 125 days after explosion, indicating that SN\,2017dwh produced a lower mass of nickel \citep[SN\,2011bm produced $\approx\!0.7$ M$_{\odot}$;][]{Valenti2012}.  Reconciling this with the much higher luminosity of SN\,2017dwh at peak requires a different power mechanism for the peak of the light curve.  In addition, we find that the upper limit on the luminosity of SN\,2017dwh at $\approx\!300$ days after explosion falls just above the tail luminosities of SN\,1998bw and SN\,2004aw.  As found for the SLSN-I PS16aqv \citep{Blanchard2018}, this implies that SN\,2017dwh did not produce significantly more nickel than that produced in Type Ic/Ic-BL SNe.  Our limits leave open the possibility that SN\,2017dwh produced roughly a similar amount of nickel as these SNe \citep[SN\,1998bw produced $\approx\!0.4$ M$_\odot$ of $^{56}$Ni;][]{Sollerman2002}.  Under similar assumptions for the gamma-ray optical depth evolution \citep{Sollerman2002}, we estimate an upper limit on the nickel mass in SN\,2017dwh of $M_{\rm Ni} \lesssim 0.6$ M$_{\odot}$.   

\begin{figure*}
\centering
\includegraphics[scale=0.8]{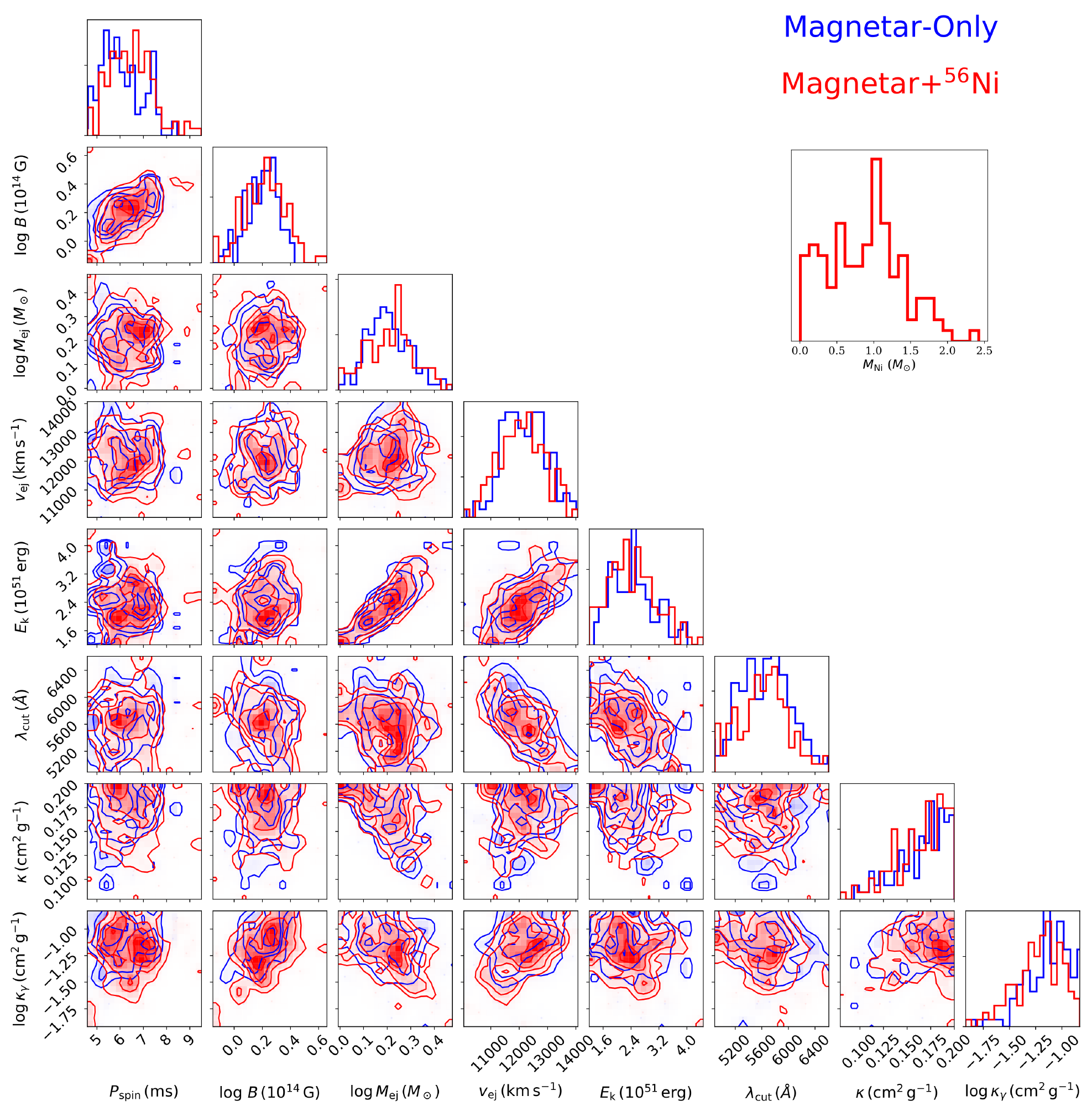}
\caption{Corner plot showing the posterior parameter distributions for the magnetar-only (blue) and the combined magnetar and $^{56}$Ni (red) model fits with {\tt MOSFiT}.  The posterior on $M_{\rm Ni}$ for the latter model is shown separately. The model light curves corresponding to the magnetar-only model are compared to the observed data in Figure \ref{fig:mosfit}.  We find that the distributions of these parameters overlap between the two models.  In other words, adding energy input from the decay of $^{56}$Ni does not significantly impact the magnetar and ejecta parameters, indicating that the dominating effect of the magnetar can mask energy input from radioactive decay even with large $^{56}$Ni masses of $\approx\!1.0$\,M$_\odot$.  Corner plot made using {\tt corner.py} \citep{corner2016}.}
\label{fig:corner}
\end{figure*}

\subsection{Light Curve Modeling}
\label{modeling}
As for previous SLSNe-I \citep[e.g.,][]{Pastorello2010,Inserra2013}, the nickel mass implied by the late-time limits is far below that required to power the peak luminosity of SN\,2017dwh.  Consequently the required additional energy input to explain the luminous light curves makes it difficult to constrain the contribution from radioactive decay.  To quantify this, we examine how much nickel can be ``hidden" in the ejecta, assuming that the light curve peak is powered by magnetar spin-down.

\begin{deluxetable}{ccc}[h!]
\tablecolumns{3}
\tabcolsep0.1in\footnotesize
\tablewidth{4in}
\tablecaption{Model parameter medians and $\pm1\sigma$ ranges corresponding to the posteriors in Figure \ref{fig:corner} associated with the magnetar-only (realizations shown in Figure \ref{fig:mosfit}) and combined magnetar and $^{56}$Ni fits   
\label{tab:param}}
\tablehead {
\colhead {Parameter}   &
\colhead {Magnetar-only} &
\colhead {Magnetar+$^{56}$Ni}
}   
\startdata
$P_{\rm spin}$ (ms) & 5.93$^{+1.33}_{-0.70}$ &  6.41$^{+0.89}_{-0.89}$ \\[5pt]
log($B$/10$^{14}$ G) & $0.21^{+0.10}_{-0.13}$ & 0.22$^{+0.14}_{-0.16}$ \\[5pt]
log($M_{\rm ej}$/M$_{\odot}$) & 0.20$^{+0.10}_{-0.08}$ & 0.22$^{+0.08}_{-0.12}$ \\[5pt] 
$f_{\rm Ni}$ & \nodata & 0.57$^{+0.28}_{-0.38}$ \\[5pt]
$M_{\rm Ni}$ (M$_{\odot}$) & \nodata & 0.89$^{+0.52}_{-0.58}$ \\[5pt]
$v_{\rm ej}$ (km s$^{-1}$) & 12200$^{+800}_{-700}$ & 12100$^{+800}_{-900}$ \\[5pt] 
$E_{\rm k}$ (10$^{51}$ erg) & 2.44$^{+0.72}_{-0.61}$ & 2.36$^{+0.72}_{-0.61}$ \\[5pt]
$\lambda_{\rm cut}$ (\AA) & 5650$^{+370}_{-420}$ & 5650$^{+310}_{-440}$ \\[5pt]
$\kappa$ (cm$^{2}$ g$^{-1}$) & 0.17$^{+0.02}_{-0.03}$ & 0.17$^{+0.02}_{-0.04}$ \\[5pt]
log $\!\kappa_{\gamma}$ & $-1.09^{+0.24}_{-0.25}$ & $-1.19^{+0.19}_{-0.26}$ \\[5pt]
$M_{\rm NS}$ (M$_{\odot}$) & 1.81$^{+0.25}_{-0.23}$ & 1.82$^{+0.29}_{-0.26}$ \\[5pt]
$T_{\rm min}$ (K) & 4650$^{+710}_{-320}$ & 4550$^{+440}_{-220}$ \\[5pt]
$A_{\rm V}^{\rm host}$ & 0.39$^{+0.08}_{-0.17}$ & 0.40$^{+0.07}_{-0.14}$ \\[5pt]
$t_{\rm exp}$ (days) & $-4.16^{+0.62}_{-0.77}$ & $-4.49^{+0.93}_{-0.84}$ \\[5pt] 
log $\!\sigma$ & $-0.96^{+0.05}_{-0.07}$ & $-0.97^{+0.08}_{-0.07}$
\enddata
\tablecomments{$P_{\rm spin}$ is the initial spin period of the magnetar, $B$ is the component of the magnetar magnetic field perpendicular to the spin axis, $M_{\rm ej}$ is the ejecta mass, $f_{\rm Ni}$ is the fraction of the ejecta mass composed of $^{56}$Ni, $M_{\rm Ni}$ is the mass of $^{56}$Ni, $v_{\rm ej}$ is the ejecta velocity, $E_{\rm k}$ is the kinetic energy, $\lambda_{\rm cut}$ is the blackbody cutoff wavelength used to account for UV line-blanketing, $\kappa$ is the opacity, $\!\kappa_{\gamma}$ is the gamma-ray opacity, $M_{\rm NS}$ is the neutron star mass, $T_{\rm min}$ is the photosphere temperature floor, $A_{\rm V}^{\rm host}$ is the internal host galaxy extinction, $t_{\rm exp}$ is the explosion time relative to the first observation, and $\!\sigma$ is the uncertainty required to yield a reduced chi-squared of 1.  For more details on the model and these parameters see \citet{Nicholl2017}.} 
\end{deluxetable}

We use the Modular Open Source Fitter for Transients \citep[{\tt MOSFiT};][]{Guillochon2018}, an open source modular transient Markov Chain Monte Carlo (MCMC) code, to fit the multi-band observed light curves of SN\,2017dwh.  We fit the SN with two models, one where the energy input is solely provided by a magnetar central engine and another in which energy from the radioactive decay of $^{56}$Ni is also included.  We use the same base SLSN-I magnetar model and priors as those used in \citet{Nicholl2017}, except for one modification.  In \citet{Nicholl2017} the blackbody cut-off wavelength ($\lambda_{\rm cut}$; used to account for line blanketing in the UV) is fixed at 3000 \AA.  Due to the significant absorption from Fe-group elements in SN\,2017dwh, the appropriate cut-off wavelength is shifted redward and so we add this as a free parameter with a flat prior at $\lambda_{\rm cut} = 2500-7500$ \AA.  For the combined magnetar and $^{56}$Ni model we parameterize the nickel mass as a fraction $f_{\rm Ni}$ of the total ejecta mass, and use a flat prior of $f_{\rm Ni} = 0 - 1$.   We ran the MCMC fitting procedure until convergence was reached as assessed using the condition that the Potential Scale Reduction Factor is $<1.1$ \citep{GelmanRubin1992,BrooksGelman1998}.  This typically equates to about $25,000-40,000$ iterations.     

We overlay the ensemble of model realizations for the magnetar-only model on the observed light curves in Figure \ref{fig:mosfit}.  We find a good fit to the data (the realizations with $^{56}$Ni appear indistinguishable).  We show the parameter posterior distributions for both the magnetar-only and combined magnetar and $^{56}$Ni models in Figure \ref{fig:corner} and report the median parameter values and their $\pm1\sigma$ ranges in Table \ref{tab:param}.  We find that there is no significant difference in any parameter between the magnetar-only and the combined magnetar and $^{56}$Ni model fits.  

The posterior distribution for $f_{\rm Ni}$ is quite broad with multiple peaks and a median value of 0.57$^{+0.28}_{-0.38}$, indicating $f_{\rm Ni}$ is not well constrained.  The corresponding combined posterior on $M_{\rm Ni} = f_{\rm Ni}M_{\rm ej}$ (shown in Figure \ref{fig:corner}) yields a median value of 0.89$_{-0.58}^{+0.52}$ M$_{\odot}$.  The implied 90\% confidence upper limit on the nickel mass is $M_{\rm Ni} \lesssim 1.5$ M$_\odot$. The combined magnetar and $^{56}$Ni model can accommodate both a large and small mass of nickel due to the dominating effect of the magnetar power required to explain the peak luminosity.  We therefore conclude that it is possible for a magnetar-powered light curve to mask even considerable radioactive power input. Importantly, the model can easily accommodate a nickel mass comparable to GRB-SNe \citep[e.g., SN\,1998bw had $M_{\rm Ni} \approx 0.3-0.7$ M$_{\odot}$;][]{Iwamoto1998,Nakamura2001,Sollerman2002}.  The upper limit on $M_{\rm Ni}$ inferred from our late-time observations ($M_{\rm Ni} \lesssim 0.6$ M$_\odot$) falls within the $\pm1\sigma$ range of model-derived nickel masses, though the model can accommodate slightly higher nickel masses due to the dominating influence of the early-time data on the inferred parameters.  

We also note that we find rather atypical magnetar and ejecta properties for SN\,2017dwh compared to the large sample of SLSNe-I in \citet{Nicholl2017}.  The initial spin period of $\approx\!6$ ms is one of the longest inferred values, well outside the 1$\sigma$ range of $1.2-4$ ms for the distribution.  The ejecta mass of $\approx\!1.6$ M$_{\odot}$ is one of the lowest values, outside the 1$\sigma$ range of $2.2-12.9$ M$_{\odot}$ for the distribution.

\section{Discussion} \label{sec:disc}
We have shown that the spectroscopic features and evolution exhibited by SN\,2017dwh are unusual in the sample of SLSNe-I.  While being as luminous as SLSNe-I and bluer than Type Ic/Ic-BL SNe at peak, the SN exhibited an unusually strong line we identify as \ion{Co}{2} and rapidly evolved to closely match the much redder spectra of Type Ic/Ic-BL SNe.  By modeling and comparing the light curve to Type Ic/Ic-BL SNe, we find that it can accommodate a nickel mass similar to GRB-SNe ($\approx\!0.5$ M$_{\odot}$).

\subsection{A nickel-rich SLSN}

While most SLSNe-I have poor constraints on the mass of synthesized $^{56}$Ni, deep limits on the late-time luminosity of PS16aqv \citep{Blanchard2018} suggest that the mass of $^{56}$Ni produced in at least some SLSNe-I is not much more than the masses inferred for Type Ic/Ic-BL SNe. Our observations of SN\,2017dwh are consistent with a magnetar-powered SN that exhibited a higher-than-usual fraction of synthesized $^{56}$Ni.  In particular, the strong line from \ion{Co}{2}, similar to that seen in Type Ia SNe, is indicative of a high fraction of $^{56}$Ni in the ejecta of SN\,2017dwh.  We find that the light curve leaves open the possibility that SN\,2017dwh produced a similar amount of $^{56}$Ni as that produced in energetic Type Ic-BL SNe.  

A Type Ic-BL-like nickel mass is also consistent with the post-peak spectral evolution where SN\,2017dwh strongly resembles the red spectra of Type Ic-BL SNe, as well as the moderately energetic Type Ic SN\,2004aw.  However, due to the energy input from the magnetar central engine, at peak SN\,2017dwh is bluer than Type Ic/Ic-BL SNe and appears more similar to SLSNe-I.  

In addition, the relatively low ejecta mass of 1.6 M$_{\odot}$ from light curve modeling also supports a high $^{56}$Ni fraction.  Using the upper limit of $M_{\rm Ni} \lesssim 0.6$ M$_{\odot}$ from our light curve comparisons, the inferred ejecta mass implies a $^{56}$Ni fraction that could be as high as 30\%.  In summary, SN\,2017dwh is a transitional event with a Ic-BL-like Fe-group abundance but SLSN-like luminosity.  

By comparing SN\,2017dwh to other SLSNe-I in the literature with noticeable absorption near the \ion{Co}{2} line, we find that there may be a continuum of Fe-group absorption strength in SLSNe-I, where SN\,2017dwh is the most extreme example of high Fe-group element abundance.  The correlation between the \ion{Co}{2} absorption strength with continuum shape (Figure \ref{fig:SLSNcomp}) suggests that at least some of the variation in SLSN-I spectral shapes and evolution is due to variation in Fe-group abundance (in addition to variation in magnetar energy input).   

To assess the possibility that the magnetar/explosion properties (namely the long initial spin and low ejecta mass) are connected to the unusual spectroscopic properties, we examined several SLSNe-I with similar inferred parameters.  In the sample of SLSNe-I in \citet{Nicholl2017}, PS1-10bzj \citep{Lunnan2013} has the most similar parameters to SN\,2017dwh with $P_{\rm spin} = 5.21$ ms, log($B/10^{14}$ G) = 0.21, and $M_{\rm ej} = 1.65$ M$_{\odot}$.  However, PS1-10bzj does not show evidence for a strong absorption line due to \ion{Co}{2} nor does it exhibit rapid reddening.  Another event with a similarly long initial spin period and low ejecta mass is DES13S2cmm \citep{Papadopoulos2015}.  While the one spectrum of this event does not cover \ion{Co}{2}, its spectral shape is consistent with most SLSNe-I for the given phase and is not unusually red.  While the number of events falling in this region of parameter space is fairly small, the lack of evidence for enhanced Fe-group absorption in PS1-10bzj and DES13S2cmm may indicate that such absorption in SN\,2017dwh is not related to the magnetar parameters or ejecta mass.  In addition, the event with the most similar spectroscopic features to SN\,2017dwh, PTF11rks, does not have similar magnetar parameters or ejecta mass ($P_{\rm spin} = 2.07$ ms, log($B/10^{14}$ G) = 0.46, $M_{\rm ej} = 6.54$ M$_{\odot}$), though we note this event had one of the least-constrained fits in the sample of \citet{Nicholl2017}.   

\subsection{A new link between SLSNe-I and Type Ic-BL SNe}

SN\,2017dwh provides new evidence supporting the picture where the progenitors of SLSNe-I are similar to those of Type Ic-BL SNe.  Previously, the strongest links have come from (i) comparing the nebular spectra of SLSNe-I with Type Ic-BL SNe with and without LGRBs which suggests the progenitors have a similar inner structure/composition \citep{Milisavljevic2013,Nicholl2016,Jerkstrand2016,Jerkstrand2017,Nicholl2018}, (ii) the similar host galaxies of SLSNe-I and LGRBs which implies that they require similar environmental conditions \citep{Chen2013,Lunnan2014,Leloudas2015,Perley2016,Schulze2018}, and (iii) the luminous SN\,2011kl associated with an ultra-long GRB \citep{Greiner2015}.  Adding to these connections, SN\,2017dwh provides the strongest link yet between the photospheric phase spectra of SLSNe-I and Type Ic-BL SNe, suggesting that at least some SLSNe-I have {\it outer ejecta} compositions similar to Type Ic-BL SNe.  SN\,2017dwh can be viewed as a nickel-rich SLSN-I that transitioned to appear like a Type Ic-BL SNe.  

However, the key questions are what physical factors make SN\,2017dwh transitional, why SLSNe-I such as SN\,2017dwh are so rare in the sample, and how a significant amount of Fe-group elements reach the outer layers of the ejecta.  The answers may naturally be a consequence of the central engine, since engines can produce aspherical explosions \citep{Maeda2002}.  The formation of a jet by a magnetar central engine \citep{Metzger2011}, which may or may not successfully escape the star, may assist with the outward mixing of Fe-group elements.  Nebular phase spectra of core-collapse SNe show evidence for asphericity in their inner ejecta, with GRB-SNe being the most aspherical \citep{Maeda2008}.  While we do not have nebular spectra of SN\,2017dwh, we note that some SLSNe-I have shown asymmetric nebular line profiles \citep{Nicholl2018}.  Synthetic photospheric spectra of SNe resulting from jet-driven explosions are redder when viewed along the polar direction than the equatorial direction, showing that asphericity likely affects the spectral shape and line widths of a SN during the photospheric phase \citep{Barnes2018}.  A viewing angle close to the jet axis has been suggested as a possible explanation for high-velocity Fe-group features in the spectra of GRB-SNe such as SN\,1998bw \citep{Maeda2002} and SN\,2016jca \citep{Ashall2017}.  A combination of jet formation success rate, the related degree of asphericity of SLSNe-I, and viewing angle may account for the rarity of such strong Fe-group absorption in SLSNe-I.     

\subsection{The low-luminosity host galaxy}
\label{sec:host}

Finally, we note that SN\,2017dwh occurred in a very faint galaxy.  In Section \ref{sec:LC} we described our late-time observations which revealed a non-varying source at the position of SN\,2017dwh.  We find the following magnitudes for this source (corrected for Galactic extinction): $m_{g} = 25.27 \pm 0.10$, $m_{r} = 25.45 \pm 0.16$, and $m_{i} = 25.30 \pm 0.14$.  Using color transformations from \citet{Jordi2006} we find an absolute $B$-band luminosity of $M_{B} = -13.5$ mag, indicating that the host galaxy of SN\,2017dwh is the faintest detected SLSN-I host galaxy \citep{Lunnan2014,Leloudas2015,Perley2016,Schulze2018}.  For comparison, the $z<0.5$ host sample from \citet{Schulze2018} has a mean luminosity of $M_{B} = -17.1$ mag with a sample standard deviation of 1.5 mag and the host of SN\,2017dwh is fainter than the next faintest galaxy in this sample by 0.9 mag.  

Using the luminosity-metallicity relationship for dwarf galaxies found by \citet{Lee2006} we estimate a metallicity of $12 + {\rm log(O/H)} \sim 7.6 \sim 0.08\,Z_{\odot}$ \citep{Asplund2009}.  One might expect low metallicity to be reflected in the spectra of the SN, rather than the high Fe-group abundance we find here.  However, $^{56}$Ni production is more sensitive to progenitor star structure and explosion dynamics than to metallicity.  Whether the environmental properties of SN\,2017dwh's host galaxy are linked to its peculiar spectral properties is difficult to assess with a single event, though we note that sample studies have not confirmed correlations between host galaxy and magnetar/explosion properties \citep{Nicholl2017,DeCia2018} first suggested by \citet{Chen2017}\footnote{\citet{Chen2017} suggested that SLSNe-I with the fastest magnetar spin periods are found in the most metal-poor galaxies, which is the opposite of what our findings imply.}.  However, it is suggestive that such an unusual SLSN-I occurred in such an extreme environment.       

\section{Summary and Conclusions}
We presented photometric and spectroscopic observations of the unusual SLSN-I SN\,2017dwh.  We summarize the key findings below:

\begin{itemize}
\item The near-peak spectra of SN\,2017dwh exhibited an unusually strong absorption feature centered near 3200 \AA.  Through comparison with synthetic spectra generated with {\tt SYN++}, we identify \ion{Co}{2} as a major contributor to this feature.  These spectra also exhibit a redder continuum than typical SLSNe-I.

\item Comparing the spectra with previous SLSNe-I we find a correlation between continuum shape and \ion{Co}{2} absorption strength where the reddest events exhibit the strongest absorption.  SN\,2017dwh exhibits the most extreme example of strong absorption and red continuum. 

\item Analyzing the spectral evolution, we find that SN\,2017dwh more closely matches SLSNe-I near peak but then rapidly evolves to match the red spectra of Type Ic/Ic-BL SNe about a month after peak.

\item The spectra suggest a high fraction of synthesized $^{56}$Ni in the ejecta of SN\,2017dwh possibly due to enhanced $^{56}$Ni production or more efficient outward mixing of Fe-group elements.

\item By comparing and modeling the light curve, we are unable to rule out a synthesized $^{56}$Ni mass as high as that observed in GRB-SNe (we estimate $M_{\rm Ni} \lesssim0.6$ M$_{\odot}$).  We find a low ejecta mass, which may aid in having a high $^{56}$Ni fraction relative to other SLSNe-I if $^{56}$Ni production and ejecta mass are largely decoupled.  

\item SN\,2017dwh occurred in the faintest detected host galaxy for a SLSN-I, with $M_{B} = -13.5$ and an implied metallicity of $12 + {\rm log(O/H)} \sim 7.6 \sim 0.08\,Z_{\odot}$.
\end{itemize}

SN\,2017dwh provides a direct link to lower luminosity Type Ic/Ic-BL SNe, in the {\it photospheric} phase.  While spectral similarities between SLSNe-I and Type Ic/Ic-BL SNe have been noted before, SN\,2017dwh is unprecedented among SLSNe-I in its rapid spectroscopic evolution to match the red colors of Type Ic/Ic-BL SNe.  The strong \ion{Co}{2} absorption feature further sets it apart from typical SLSNe-I.  Although constraining the absolute mass of $^{56}$Ni produced in SLSNe-I is difficult because another energy source is dominating the luminosity, SN\,2017dwh clearly shows a greater impact of Fe-group elements in its spectra than seen in previous SLSNe-I.  The rarity of such events in the population may be the result of an unusually efficient mixing process in some events, perhaps due to a jet-like outflow.  SN\,2017dwh is a transitional event which provides new evidence supporting the picture previously inferred from host galaxies and nebular spectra.  Finally, the correlation between continuum shape and \ion{Co}{2} absorption strength suggests that variation in Fe-group abundance is responsible for some of the variation in the spectral shapes of SLSNe-I.  

\acknowledgments
The Berger Time-Domain Group at Harvard is supported in part by the NSF under grant AST-1714498 and by NASA under grant NNX15AE50G.  This paper is based upon work supported by the National Science Foundation Graduate Research Fellowship Program under Grant No. DGE1144152.  This work is based in part on observations obtained at the MDM Observatory, operated by Dartmouth College, Columbia University, Ohio State University, Ohio University, and the University of Michigan.  This paper uses data products produced by the OIR Telescope Data Center, supported by the Smithsonian Astrophysical Observatory.  Some observations reported here were obtained at the MMT Observatory, a joint facility of the Smithsonian Institution and the University of Arizona.  This paper includes data gathered with the 6.5 meter Magellan Telescopes located at Las Campanas Observatory, Chile.  This paper made use of the Open Supernova Catalog \citep{Guillochon2017}.

\appendix

\section{Observation Information and Photometry}

We present the details of the spectroscopic observations in Table \ref{tab:spec} and the photometry in Table \ref{tab:phot}.

\begin{deluxetable*}{ccccccc}[!htb]
\tablecolumns{7}
\tabcolsep0.1in\footnotesize
\tablewidth{7in}
\tablecaption{Spectroscopic Observations of SN\,2017dwh  
\label{tab:spec}}
\tablehead {
\colhead {Date}   &
\colhead {MJD}     &
\colhead {Phase\tablenotemark{a}} &
\colhead {Telescope} &
\colhead{Instrument}  &
\colhead {Airmass}   &
\colhead {Resolution (\AA)}           
}   
\startdata
19 May 2017 & 57893.4 & +3  & FLWO 60-inch & FAST & 1.48 & 5.7 \\
22 May 2017 & 57896.8 & +6  & MMT & Blue Channel & 1.20 & 4 \\
15 June 2017 & 57919.8 & +26  & MDM/Hiltner & OSMOS & 1.45 & 5 \\
28 June 2017 & 57933.7 & +39 & MMT & Blue Channel & 1.19 & 4 \\
17 July 2017 & 57952.5 & +55  & Magellan/Baade & IMACS & 2.07 & 5.4
\enddata
\tablenotetext{a}{Rest-frame days since $i$-band maximum}
\end{deluxetable*}

\begin{deluxetable*}{ccccccccc}
\tablecolumns{9}
\tabcolsep0.1in\footnotesize
\tablewidth{7in}
\tablecaption{Photometry of SN\,2017dwh ($grciz$ are in AB magnitudes and $UB$ and CSS are in Vega magnitudes) 
\label{tab:phot}}
\tablehead {
\colhead {MJD}   &
\colhead {$U$}     &
\colhead {$B$} &
\colhead {$g$} &
\colhead{$r$}  &
\colhead {$CSS$}   &
\colhead {$c$} &
\colhead{$i$} &
\colhead{$z$}           
}   
\startdata
57865.50 & \nodata & \nodata & \nodata & \nodata & 20.39 (0.21) & \nodata & \nodata & \nodata \\
57868.50 & \nodata & \nodata & \nodata & \nodata & 19.12 (0.08) & \nodata & \nodata & \nodata \\
57870.10 & \nodata & \nodata & \nodata & \nodata & \nodata & 18.73 (0.11) & \nodata & \nodata \\
57884.47 & \nodata & \nodata & \nodata & \nodata & \nodata & \nodata & 18.03 (0.04) & \nodata \\
57889.34 & \nodata & \nodata & \nodata & \nodata & \nodata & \nodata & 18.04 (0.13) & \nodata \\
57890.39 & \nodata & \nodata & \nodata & \nodata & \nodata & \nodata & 17.95 (0.11) & \nodata \\
57892.38 & \nodata & \nodata & \nodata & \nodata & \nodata & \nodata & 17.98 (0.09) & \nodata \\
57906.37 & \nodata & \nodata & 18.91 (0.12) & 18.36 (0.08) & \nodata & \nodata & 18.26 (0.09) & \nodata \\
57907.39 & \nodata & \nodata & 18.99 (0.11) & \nodata & \nodata & \nodata & \nodata & \nodata \\
57911.20 & 19.58 (0.23) & 20.07 (0.25) & \nodata & \nodata & \nodata & \nodata & \nodata & \nodata \\
57913.40 & \nodata & \nodata & \nodata & \nodata & \nodata & \nodata & 18.53 (0.02) & \nodata \\
57914.21 & \nodata & \nodata & 19.62 (0.12) & 18.85 (0.08) & \nodata & \nodata & 18.55 (0.13) & \nodata \\
57915.22 & \nodata & \nodata & 19.80 (0.11) & 18.89 (0.07) & \nodata & \nodata & 18.60 (0.11) & \nodata \\
57916.28 & \nodata & \nodata & 19.90 (0.12) & 18.96 (0.10) & \nodata & \nodata & 18.62 (0.12) & \nodata \\
57917.23 & \nodata & \nodata & 19.89 (0.14) & 18.97 (0.09) & \nodata & \nodata & 18.68 (0.11) & \nodata \\
57919.76 & \nodata & \nodata & 20.05 (0.07) & 18.96 (0.07) & \nodata & \nodata & 18.67 (0.06) & 18.77 (0.05) \\
57920.01 & 20.85 (0.34) & \nodata & \nodata & \nodata & \nodata & \nodata & \nodata & \nodata \\
57920.32 & \nodata & \nodata & 20.08 (0.13) & 19.00 (0.12) & \nodata & \nodata & 18.75 (0.10) & \nodata \\
57922.32 & \nodata & \nodata & 20.08 (0.11) & 19.10 (0.10) & \nodata & \nodata & 18.76 (0.10) & \nodata \\
57922.77 & \nodata & \nodata & 20.15 (0.09) & 18.99 (0.04) & \nodata & \nodata & 18.78 (0.09) & 18.82 (0.07) \\
57923.32 & \nodata & \nodata & 20.18 (0.13) & 19.04 (0.10) & \nodata & \nodata & 18.74 (0.14) & \nodata \\
57924.54 & $>21.4$ & \nodata & \nodata & \nodata & \nodata & \nodata & \nodata & \nodata \\
57934.17 & \nodata & \nodata & 20.70 (0.14) & 19.59 (0.08) & \nodata & \nodata & 19.28 (0.11) & \nodata \\
57939.54 & $>20.4$ & \nodata & \nodata & \nodata & \nodata & \nodata & \nodata & \nodata \\
57941.25 & \nodata & \nodata & \nodata & 20.07 (0.15) & \nodata & \nodata & 19.50 (0.14) & \nodata \\
57943.53 & $>20.7$ & \nodata & \nodata & \nodata & \nodata & \nodata & \nodata & \nodata \\
57952.48 & \nodata & \nodata & 22.00 (0.06) & 20.67 (0.03) & \nodata & \nodata & 20.21 (0.06) & \nodata \\
58002.62 & \nodata & \nodata & \nodata & 21.92 (0.24) & \nodata & \nodata & 21.61 (0.26) & \nodata \\
58007.10 & \nodata & \nodata & \nodata & 22.27 (0.18) & \nodata & \nodata & \nodata & \nodata \\
58009.10 & \nodata & \nodata & \nodata & $>22.3$ & \nodata & \nodata & \nodata & \nodata \\
58103.51 & \nodata & \nodata & $>24.0$ & $>23.5$ & \nodata & \nodata & \nodata & \nodata \\
58104.01 & \nodata & \nodata & \nodata & \nodata & \nodata & \nodata & $>22.7$ & \nodata \\
58156.48 & \nodata & \nodata & \nodata & \nodata & \nodata & \nodata & $>24.1$ & \nodata \\
58195.79 & \nodata & \nodata & \nodata & $>25.7$ & \nodata & \nodata & \nodata & \nodata \\
\enddata
\tablecomments{These magnitudes are not corrected for Galactic extinction.}
\end{deluxetable*}


\begin{thebibliography}{}

 

\bibitem[Ashall et al.(2017)]{Ashall2017} Ashall, C., Pian, E., Mazzali, P.~A., et al.\ 2017, ArXiv e-prints , arXiv:1702.04339

\bibitem[Asplund et al.(2009)]{Asplund2009} Asplund, M., Grevesse, N., Sauval, A.~J., et al.\ 2009, Annual Review of Astronomy and Astrophysics, 47, 481

\bibitem[Barnes et al.(2018)]{Barnes2018} Barnes, J., Duffell, P.~C., Liu, Y., et al.\ 2018, \apj, 860, 38

\bibitem[Becker(2015)]{Becker2015} Becker, A.\ 2015, Astrophysics Source Code Library , ascl:1504.004

\bibitem[Blanchard et al.(2018)]{Blanchard2018} Blanchard, P.~K., Nicholl, M., Berger, E., et al.\ 2018, \apj, 865, 9

\bibitem[Breeveld et al.(2011)]{Breeveld11} Breeveld, A.~A., Landsman, W., Holland, S.~T., Roming, P., Kuin, N.~P.~M.,  \& Page, M.~J.\ 2011, in American Institute of Physics Conference Series, Vol. 1358, American Institute of Physics Conference Series, ed. J.~E. McEnery, J.~L. Racusin, \& N.~Gehrels, 373

\bibitem[Brooks \& Gelman(1998)]{BrooksGelman1998} Brooks, S.~P. \& Gelman, A.\ 1998, Journal of computational and graphical statistics, 7, 434

\bibitem[Brown et al.(2009)]{Brown09} Brown, P.~J., et al.\ 2009, \aj, 137, 4517

\bibitem[Cano et al.(2017)]{Cano2017} Cano, Z., Izzo, L., de Ugarte Postigo, A., et al.\ 2017, \aap, 605, A107

\bibitem[Chatzopoulos et al.(2013)]{Chatzopoulos2013} Chatzopoulos, E., Wheeler, J.~C., Vinko, J., et al.\ 2013, \apj, 773, 76

\bibitem[Chen et al.(2013)]{Chen2013} Chen, T.-W., Smartt, S.~J., Bresolin, F., et al.\ 2013, \apj, 763, L28

\bibitem[Chen et al.(2017)]{Chen2017} Chen, T.-W., Smartt, S.~J., Yates, R.~M., et al.\ 2017, \mnras, 470, 3566

\bibitem[Chevalier \& Irwin(2011)]{ChevalierIrwin2011} Chevalier, R.~A. \& Irwin, C.~M.\ 2011, \apj, 729, L6

\bibitem[Chomiuk et al.(2011)]{Chomiuk2011} Chomiuk, L., Chornock, R., Soderberg, A.~M., et al.\ 2011, \apj, 743, 114

\bibitem[De Cia et al.(2018)]{DeCia2018} De Cia, A., Gal-Yam, A., Rubin, A., et al.\ 2018, \apj, 860, 100

\bibitem[Dessart et al.(2012)]{Dessart2012} Dessart, L., Hillier, D.~J., Waldman, R., et al.\ 2012, \mnras, 426, L76

\bibitem[Drake et al.(2009)]{Drake2009} Drake, A.~J., Djorgovski, S.~G., Mahabal, A., et al.\ 2009, \apj, 696, 870

\bibitem[Dressler et al.(2011)]{Dressler2011} Dressler, A., Bigelow, B., Hare, T., et al.\ 2011, Publications of the Astronomical Society of the Pacific, 123, 288

\bibitem[Fabricant et al.(1998)]{Fabricant1998} Fabricant, D., Cheimets, P., Caldwell, N., et al.\ 1998, Publications of the Astronomical Society of the Pacific, 110, 79

\bibitem[Filippenko et al.(1992)]{Filippenko1992} Filippenko, A.~V., Richmond, M.~W., Matheson, T., et al.\ 1992, \apj, 384, L15

\bibitem[Foley et al.(2003)]{Foley2003} Foley, R.~J., Papenkova, M.~S., Swift, B.~J., et al.\ 2003, Publications of the Astronomical Society of the Pacific, 115, 1220

\bibitem[Foreman-Mackey(2016)]{corner2016} Foreman-Mackey, D.\ 2016, The Journal of Open Source Software, 1, 24

\bibitem[Galama et al.(1998)]{Galama1998} Galama, T.~J., Vreeswijk, P.~M., van Paradijs, J., et al.\ 1998, \nat, 395, 670

\bibitem[Gal-Yam et al.(2002)]{Gal-Yam2002} Gal-Yam, A., Ofek, E.~O., \& Shemmer, O.\ 2002, \mnras, 332, L73

\bibitem[Gal-Yam et al.(2009)]{Gal-Yam2009} Gal-Yam, A., Mazzali, P., Ofek, E.~O., et al.\ 2009, \nat, 462, 624

\bibitem[Gal-Yam(2012)]{Gal-Yam2012} Gal-Yam, A.\ 2012, Science, 337, 927

\bibitem[Gelman \& Rubin(1992)]{GelmanRubin1992} Gelman, A. \& Rubin, D.~B.\ 1992, Statistical Science, 7, 457

\bibitem[Greiner et al.(2015)]{Greiner2015} Greiner, J., Mazzali, P.~A., Kann, D.~A., et al.\ 2015, \nat, 523, 189

\bibitem[Guillochon et al.(2017)]{Guillochon2017} Guillochon, J., Parrent, J., Kelley, L.~Z., et al.\ 2017, \apj, 835, 64

\bibitem[Guillochon et al.(2018)]{Guillochon2018} Guillochon, J., Nicholl, M., Villar, V.~A., et al.\ 2018, The Astrophysical Journal Supplement Series, 236, 6

\bibitem[Heger \& Woosley(2002)]{HegerWoosley2002} Heger, A. \& Woosley, S.~E.\ 2002, \apj, 567, 532

\bibitem[Huber et al.(2015)]{Huber2015} Huber, M., Chambers, K.~C., Flewelling, H., et al.\ 2015, The Astronomer's Telegram, 7153

\bibitem[Inserra et al.(2013)]{Inserra2013} Inserra, C., Smartt, S.~J., Jerkstrand, A., et al.\ 2013, \apj, 770, 128

\bibitem[Iwamoto et al.(1998)]{Iwamoto1998} Iwamoto, K., Mazzali, P.~A., Nomoto, K., et al.\ 1998, \nat, 395, 672

\bibitem[Iwamoto et al.(2000)]{Iwamoto2000} Iwamoto, K., Nakamura, T., Nomoto, K., et al.\ 2000, \apj, 534, 660

\bibitem[Jerkstrand et al.(2016)]{Jerkstrand2016} Jerkstrand, A., Smartt, S.~J. \& Heger, A.\ 2016, \mnras, 455, 3207

\bibitem[Jerkstrand et al.(2017)]{Jerkstrand2017} Jerkstrand, A., Smartt, S.~J., Inserra, C., et al.\ 2017, \apj, 835, 13

\bibitem[Jordi et al.(2006)]{Jordi2006} Jordi, K., Grebel, E.~K. \& Ammon, K.\ 2006, \aap, 460, 339

\bibitem[Kasen \& Bildsten(2010)]{KasenBildsten2010} Kasen, D., \& Bildsten, L.\ 2010, \apj, 717, 245

\bibitem[Lee et al.(2006)]{Lee2006} Lee, H., Skillman, E.~D., Cannon, J.~M., et al.\ 2006, \apj, 647, 970

\bibitem[Leloudas et al.(2015)]{Leloudas2015} Leloudas, G., Schulze, S., Kr{\"u}hler, T., et al.\ 2015, \mnras, 449, 917

\bibitem[Lunnan et al.(2013)]{Lunnan2013} Lunnan, R., Chornock, R., Berger, E., et al.\ 2013, \apj, 771, 97

\bibitem[Lunnan et al.(2014)]{Lunnan2014} Lunnan, R., Chornock, R., Berger, E., et al.\ 2014, \apj, 787, 138

\bibitem[MacFadyen, \& Woosley(1999)]{MacFadyen1999} MacFadyen, A.~I., \& Woosley, S.~E.\ 1999, \apj, 524, 262

\bibitem[Maeda et al.(2002)]{Maeda2002} Maeda, K., Nakamura, T., Nomoto, K., et al.\ 2002, \apj, 565, 405

\bibitem[Maeda et al.(2008)]{Maeda2008} Maeda, K., Kawabata, K., Mazzali, P.~A., et al.\ 2008, Science, 319, 1220

\bibitem[Margalit et al.(2018)]{Margalit2018} Margalit, B., Metzger, B.~D., Thompson, T.~A., et al.\ 2018, \mnras, 475, 2659

\bibitem[Martini et al.(2011)]{Martini2011} Martini, P., Stoll, R., Derwent, M.~A., et al.\ 2011, Publications of the Astronomical Society of the Pacific, 123, 187

\bibitem[Mazzali(2000)]{Mazzali2000} Mazzali, P.~A.\ 2000, \aap, 363, 705

\bibitem[Mazzali et al.(2002)]{Mazzali2002} Mazzali, P.~A., Deng, J., Maeda, K., et al.\ 2002, \apj, 572, L61

\bibitem[Mazzali et al.(2005)]{Mazzali2005} Mazzali, P.~A., Kawabata, K.~S., Maeda, K., et al.\ 2005, Science, 308, 1284

\bibitem[Mazzali et al.(2014)]{Mazzali2014} Mazzali, P.~A., McFadyen, A.~I., Woosley, S.~E., et al.\ 2014, \mnras, 443, 67

\bibitem[Mazzali, et al.(2016)]{Mazzali2016} Mazzali, P.~A., Sullivan, M., Pian, E., et al.\ 2016, \mnras, 458, 3455

\bibitem[Mazzali et al.(2017)]{Mazzali2017} Mazzali, P.~A., Sauer, D.~N., Pian, E., et al.\ 2017, \mnras, 469, 2498

\bibitem[McKenzie, \& Schaefer(1999)]{McKenzie1999} McKenzie, E.~H., \& Schaefer, B.~E.\ 1999, Publications of the Astronomical Society of the Pacific, 111, 964

\bibitem[Metzger et al.(2011)]{Metzger2011} Metzger, B.~D., Giannios, D., Thompson, T.~A., et al.\ 2011, \mnras, 413, 2031

\bibitem[Metzger et al.(2015)]{Metzger2015} Metzger, B.~D., Margalit, B., Kasen, D., et al.\ 2015, \mnras, 454, 3311

\bibitem[Milisavljevic et al.(2013)]{Milisavljevic2013} Milisavljevic, D., Soderberg, A.~M., Margutti, R., et al.\ 2013, \apj, 770, L38

\bibitem[Nakamura et al.(2001)]{Nakamura2001} Nakamura, T., Mazzali, P.~A., Nomoto, K., et al.\ 2001, \apj, 550, 991

\bibitem[Nicholl et al.(2013)]{Nicholl2013} Nicholl, M., Smartt, S.~J., Jerkstrand, A., et al.\ 2013, \nat, 502, 346

\bibitem[Nicholl et al.(2014)]{Nicholl2014} Nicholl, M., Smartt, S.~J., Jerkstrand, A., et al.\ 2014, \mnras, 444, 2096

\bibitem[Nicholl et al.(2015)]{Nicholl2015} Nicholl, M., Smartt, S.~J., Jerkstrand, A., et al.\ 2015, \mnras, 452, 3869

\bibitem[Nicholl et al.(2016)]{Nicholl2016} Nicholl, M., Berger, E., Margutti, R., et al.\ 2016, \apj, 828, L18

\bibitem[Nicholl et al.(2017)]{Nicholl2017} Nicholl, M., Guillochon, J., \& Berger, E.\ 2017, \apj, 850, 55

\bibitem[Nicholl et al.(2018)]{Nicholl2018} Nicholl, M., Berger, E., Blanchard, P.~K., et al.\ 2018, ArXiv e-prints , arXiv:1808.00510

\bibitem[Papadopoulos et al.(2015)]{Papadopoulos2015} Papadopoulos, A., D'Andrea, C.~B., Sullivan, M., et al.\ 2015, \mnras, 449, 1215

\bibitem[Pastorello et al.(2010)]{Pastorello2010} Pastorello, A., Smartt, S.~J., Botticella, M.~T., et al.\ 2010, \apj, 724, L16

\bibitem[Patat et al.(2001)]{Patat2001} Patat, F., Cappellaro, E., Danziger, J., et al.\ 2001, \apj, 555, 900

\bibitem[Perley et al.(2016)]{Perley2016} Perley, D.~A., Quimby, R.~M., Yan, L., et al.\ 2016, \apj, 830, 13

\bibitem[Planck Collaboration et 
al.(2014)]{Planck2013} Planck Collaboration, Ade, P.~A.~R., Aghanim, N., et al.\ 2014, \aap, 571, A16

\bibitem[Quimby et al.(2011)]{Quimby2011} Quimby, R.~M., Kulkarni, S.~R., Kasliwal, M.~M., et al.\ 2011, \nat, 474, 487

\bibitem[Quimby et al.(2018)]{Quimby2018} Quimby, R.~M., De Cia, A., Gal-Yam, A., et al.\ 2018, \apj, 855, 2

\bibitem[Roming et al.(2005)]{Roming05} Roming, P.~W.~A., et~al.\ 2005, \ssr, 120, 95

\bibitem[Sasdelli et al.(2014)]{Sasdelli2014} Sasdelli, M., Mazzali, P.~A., Pian, E., et al.\ 2014, \mnras, 445, 711

\bibitem[Schlafly \& Finkbeiner(2011)]{SF2011} Schlafly, E.~F., \& Finkbeiner, D.~P.\ 2011, \apj, 737, 103

\bibitem[Schmidt et al.(1989)]{Schmidt1989} Schmidt, G.~D., Weymann, R.~J. \& Foltz, C.~B.\ 1989, Publications of the Astronomical Society of the Pacific, 101, 713

\bibitem[Schulze et al.(2018)]{Schulze2018} Schulze, S., Kr{\"u}hler, T., Leloudas, G., et al.\ 2018, \mnras, 473, 1258

\bibitem[Sollerman et al.(2002)]{Sollerman2002} Sollerman, J., Holland, S.~T., Challis, P., et al.\ 2002, \aap, 386, 944

\bibitem[Stevenson et al.(2016)]{Stevenson2016} Stevenson, K.~B., Bean, J.~L., Seifahrt, A., et al.\ 2016, \apj, 817, 141

\bibitem[Taubenberger et al.(2006)]{Taubenberger2006} Taubenberger, S., Pastorello, A., Mazzali, P.~A., et al.\ 2006, \mnras, 371, 1459

\bibitem[Thomas et al.(2011)]{Thomas2011} Thomas, R.~C., Nugent, P.~E. \& Meza, J.~C.\ 2011, Publications of the Astronomical Society of the Pacific, 123, 237

\bibitem[Tonry et al.(2018)]{Tonry2018} Tonry, J.~L., Denneau, L., Heinze, A.~N., et al.\ 2018, Publications of the Astronomical Society of the Pacific, 130, 64505

\bibitem[Valenti et al.(2012)]{Valenti2012} Valenti, S., Taubenberger, S., Pastorello, A., et al.\ 2012, \apj, 749, L28

\bibitem[Woosley(1993)]{Woosley1993} Woosley, S.~E.\ 1993, \apj, 405, 273

\bibitem[Woosley(2010)]{Woosley2010} Woosley, S.~E.\ 2010, \apj, 719, L204

\bibitem[Yan et al.(2017)]{Yan2017} Yan, L., Quimby, R., Gal-Yam, A., et al.\ 2017, \apj, 840, 57

\end{thebibliography}
\end{document}